\documentclass[nofootinbib]{revtex4}
\usepackage{graphicx}
\usepackage{dcolumn}
\usepackage{amsmath,amssymb,epsfig}
\usepackage{paralist}
\usepackage{comment}
\usepackage{graphicx}
\usepackage{multirow}
\usepackage{color,soul}
\usepackage{suffix}
\usepackage{mathtools}

\allowdisplaybreaks

\renewcommand{\vec}[1]{\boldsymbol{\mathrm{#1}}}

\begin{document}

\title{Imaging extended sources with the solar gravitational lens}

\author{Slava G. Turyshev$^{1}$, Viktor T. Toth$^2$}

\affiliation{\vskip 3pt
$^1$Jet Propulsion Laboratory, California Institute of Technology,\\
4800 Oak Grove Drive, Pasadena, CA 91109-0899, USA}

\affiliation{\vskip 3pt
$^2$Ottawa, Ontario K1N 9H5, Canada}

\date{\today}

\begin{abstract}

We investigate the optical properties of the solar gravitational lens (SGL) with respect to an extended source located at a large but finite distance from the Sun. The static, spherically symmetric gravitational field of the Sun is modeled within the first post-Newtonian approximation of the general theory of relativity. We consider the propagation of monochromatic electromagnetic (EM) waves near the Sun. We develop, based on a Mie theory, a vector theory of diffraction that accounts for the refractive properties of the solar gravitational field. The finite distance to a point source can be accounted for using a rotation of the coordinate system to align its polar axis with the axis directed from the point source to the center of the Sun, which we call the optical axis.
We determine the EM field and study the key optical properties of the SGL in all four regions formed behind the Sun by an EM wave diffracted by the solar gravity field: the shadow, geometric optics, and weak and strong interference regions. Extended sources can then be represented as collections of point sources. We present the power density of the signal received by a telescope in the image plane. Our discussion concludes with considering the implications for imaging with the SGL.

\end{abstract}


\maketitle

\section{Introduction}
\label{sec:intro}

Direct imaging of exoplanets requires significant light amplification and very high angular resolution. The reason is that exoplanets are very small, extremely dim, and very distant targets for observation. When we consider traditional astronomical instruments---telescopes and interferometers---for this purpose, we face the sobering reality of requiring prohibitively large apertures, prohibitively long baselines, or a combination of both. For instance, to capture a single-pixel image of an Earth-like exoplanet from a distance of 100 light-years, a diffraction-limited telescope with an aperture of $\sim 90$~km would be required. Optical interferometers with moderate-size telescopes and large baselines would require signal integration times of hundreds of thousands if not millions of years to achieve a reasonable signal-to-noise ratio (SNR).  Clearly, these scenarios are impractical. These challenges lead us to examine other ways that have the potential to produce high-resolution, multipixel images of such distant, small, dim targets. This is our primary motivation for the ongoing study \cite{Turyshev:2017,Turyshev-Toth:2017,Turyshev-Toth:2019} of the solar gravitational lens (SGL).

The large heliocentric distances separating us from the beginning of the SGL focal region were previously hard to contemplate. Multiple recent developments in deep space exploration technologies (for review, see \cite{KISS:2015,Alkalai-etal:2017}), along with the fact that the Voyager 1 spacecraft was able to reach heliocentric distances beyond 140 astronomical units (AU) while still transmitting valuable data, place such distances within reach. This allows us to consider practical applications of the SGL as an ``optical instrument'' that could be used for multipixel imaging and spatially resolved spectroscopy of an exoplanet \cite{Turyshev-etal:2018}. Motivated by such a unique opportunity, we recently studied the optical properties of the SGL in various conditions. Specifically, we developed a wave-optical theory of the SGL \cite{Turyshev:2017,Turyshev-Toth:2017} and discussed its key features, including the SGL's light amplification and resolution.  In \cite{Turyshev-Toth:2019} we studied the impact of the plasma in the solar corona on light propagation in the vicinity of the Sun and the extended solar atmosphere.  As a result, most of the analytical tools that are needed to model the imaging of exoplanets have became available. With these analyses concluded, we found no major obstacles for imaging and spectroscopy applications of the SGL.

The next step was  the development of realistic imaging scenarios and relevant simulations. However, we realized that a treatment of extended sources imaged by the SGL  was still missing. Most of the published wave-optical analyses of the SGL assumed that the light comes from a source positioned at an infinite distance from the Sun. In reality, an exoplanet is a small, but {\em extended} object positioned at a large, but {\em finite} distance from us. In this paper we develop a wave-optical theory of the SGL for such sources.

Our paper is organized as follows:
Section~\ref{sec:em-waves-gr+pl} introduces the SGL and presents Maxwell's field equations for the electromagnetic (EM) field on the background of the solar gravitational monopole.
Section \ref{sec:EM-field-fin-dist} introduces the problem of finding the EM field from the source at a finite distance and establishes the general principles to finding the needed solution.
Section \ref{sec:go-em-outside} discusses the EM field in the geometric optics and weak interference regions.
Section \ref{sec:IF-region} is devoted to determining the EM field in the strong interference region.
Section \ref{sec:image_form} addresses the process of image formation with the SGL for an extended source.
Appendix~\ref{sec:rad_eq_wkb} discusses an approximate solution for the radial function that relies on the Wentzel--Kramers--Brillouin (WKB) approximation. A solution for this function is derived for the case when a plane EM wave originates at a large, but finite distance from the Sun.

\section{General properties of the Solar Gravitational Lens}
\label{sec:em-waves-gr+pl}

We consider the propagation of monochromatic light originating at a source that is positioned at a large but finite distance from the Sun, and received by a detector in the focal region of the SGL. Our objective is to investigate the effect of the finite distance from the source on image formation by the SGL.

\subsection{EM waves in a static gravitational field}
\label{sec:maxwell}

We focus on solving Maxwell's equations on the background set by the solar gravitational field. Following \cite{Turyshev:2017,Turyshev-Toth:2017,Turyshev-Toth:2018-grav-shadow}, we begin with the generally covariant form of Maxwell's equations:
{}
\begin{eqnarray}
\partial_lF_{ik}+\partial_iF_{kl}+\partial_kF_{li}=0, \qquad
\frac{1}{\sqrt{-g}}\partial_k\Big(\sqrt{-g}F^{ik}\Big)=-\frac{4\pi}{c}j^i,
\label{eq:max-eqs}
\end{eqnarray}
where $g_{mn}$ is the metric tensor and $g=\det g_{mn}$ is its determinant.\footnote{The notational conventions used in this paper are the same as in \cite{Landau-Lifshitz:1988,Turyshev-Toth:2017}: Latin indices ($i,j,k,...$) are spacetime indices that run from 0 to 3. Greek indices $\alpha,\beta,...$ are spatial indices that run from 1 to 3. In case of repeated indices in products, the Einstein summation rule applies: e.g., $a_mb^m=\sum_{m=0}^3a_mb^m$.  Bold letters denote spatial (three-dimensional) vectors: e.g., ${\vec a} = (a_1, a_2, a_3), {\vec b} = (b_1, b_2, b_3)$.} To describe the SGL in the first post-Newtonian approximation, we use a static harmonic metric with the line element in spherical coordinates $(r,\theta,\phi)$ (see Fig.~\ref{fig:go}), given as:
\begin{eqnarray}
ds^2&=&u^{-2}c^2dt^2-u^2\big(dr^2+r^2(d\theta^2+\sin^2\theta d\phi^2)\big),
\label{eq:metric-gen}
\end{eqnarray}
where, to the accuracy sufficient to describe light propagation in the solar system, the quantity $u$ has the form $u=1+c^{-2}U+{\cal O}(c^{-4}),$ with $U$ being the Newtonian gravitational potential. As in  \cite{Turyshev-Toth:2017}, we focus our discussion on the largest contribution to the gravitational scattering of light, which, in the case of the Sun, is due to the gravity field produced by a static monopole. In this case, the Newtonian potential may be given by $c^{-2}U({\vec r})={r_g}/{2r}+{\cal O}(r^{-3},c^{-4}),$ where $r_g=2GM_\odot/c^2$ is the Schwarzschild radius of the Sun. Therefore, the quantity $u$ in (\ref{eq:metric-gen}) has the form
{}
\begin{eqnarray}
u(r)&=&1+\frac{r_g}{2r}+{\cal O}(r^{-3},c^{-4}).
\label{eq:pot_w_1**}
\end{eqnarray}

In the case of a static, spherically symmetric gravitational field (\ref{eq:metric-gen})--(\ref{eq:pot_w_1**}), in the absence of sources or currents, $j^k\equiv (\rho,{\vec j})=0$, solving the  field equations  (\ref{eq:max-eqs}) is straightforward. We align the polar $z$-axis of the coordinate system along the wavevector $\vec k$ of the incident wave. The resulting complete solution of Maxwell's equations, following \cite{Born-Wolf:1999,Herlt-Stephani:1976}, was developed in \cite{Turyshev-Toth:2017,Turyshev-Toth:2019} with the components of the EM field ${\vec D}=u{\vec E}$ and ${\vec { B}}=u{\vec { H}}$:
{}
\begin{align}
  \left( \begin{aligned}
{   D}_r& \\
{   B}_r& \\
  \end{aligned} \right) =&  \left( \begin{aligned}
\cos\phi \\
\sin\phi  \\
  \end{aligned} \right) \,e^{-i\omega t}\alpha(r, \theta), &
    \left( \begin{aligned}
{   D}_\theta& \\
{   B}_\theta& \\
  \end{aligned} \right) =&  \left( \begin{aligned}
\cos\phi \\
\sin\phi  \\
  \end{aligned} \right) \,e^{-i\omega t}\beta(r, \theta), &
    \left( \begin{aligned}
{   D}_\phi& \\
{   B}_\phi& \\
  \end{aligned} \right) =&  \left( \begin{aligned}
-\sin\phi \\
\cos\phi  \\
  \end{aligned} \right) \,e^{-i\omega t}\gamma(r, \theta),
  \label{eq:DB-sol00p*}
\end{align}
with the quantities $\alpha, \beta$ and $\gamma$ computed from the following expressions:
{}
\begin{eqnarray}
\alpha(r, \theta)&=&
\frac{1}{u}\Big\{\frac{\partial^2 }{\partial r^2}
\Big[\frac{r\,{\hskip -1pt}\Pi}{u}\Big]+k^2 u^4\Big[\frac{r\,{\hskip -1pt}\Pi}{u}\Big]\Big\}+{\cal O}\Big(\big(\frac{1}{u}\big)''\Big),
\label{eq:alpha*}\\
\beta(r, \theta)&=&\frac{1}{u^2r}
\frac{\partial^2 \big(r\,{\hskip -1pt}\Pi\big)}{\partial r\partial \theta}+\frac{ik\big(r\,{\hskip -1pt}\Pi\big)}{r\sin\theta},
\label{eq:beta*}\\[0pt]
\gamma(r, \theta)&=&\frac{1}{u^2r\sin\theta}
\frac{\partial \big(r\,{\hskip -1pt}\Pi\big)}{\partial r}+\frac{ik}{r}
\frac{\partial\big(r\,{\hskip -1pt}\Pi\big)}{\partial \theta},
\label{eq:gamma*}
\end{eqnarray}
with $k=\omega/c$ being the wavenumber of the monochromatic EM wave and $\Pi (r, \theta)$  is the Debye potential given as \cite{Turyshev:2017,Turyshev-Toth:2017}
{}
\begin{eqnarray}
\Pi(r, \theta)&=& \frac{E_0}{2ik^2}\frac{u}{r}\sum_{\ell=kR^\star_\odot}^\infty i^{\ell-1}\frac{2\ell+1}{\ell(\ell+1)}e^{i\sigma_\ell}
H^+_\ell(kr_g, kr)P^{(1)}_\ell(\cos\theta)-\nonumber\\
&&\hskip 10 pt -\,
\frac{E_0}{2ik^2}\frac{u}{r}\sum_{\ell=1}^\infty i^{\ell-1}\frac{2\ell+1}{\ell(\ell+1)}e^{i\sigma_\ell}
H^-_\ell(kr_g, kr)P^{(1)}_\ell(\cos\theta) +{\cal O}(r_g^2),
\label{eq:Pi-s_a+0}
\end{eqnarray}
where $H^{+}_\ell$ and $H^{-}_\ell$ are the Coulomb--Hankel functions \cite{Abramovitz-Stegun:1965} representing outgoing and incident waves, correspondingly,  $\sigma_\ell$ is known as the Coulomb phase shift (see \cite{Turyshev-Toth:2017} for details), and $P^{(1)}_\ell(\cos\theta) $ are the associated Legendre polynomials.

To derive the solution for the Debye potential $\Pi(r, \theta)$ given by (\ref{eq:Pi-s_a+0}),  we used the fully absorbing boundary conditions that account for the physical size of the Sun (see details in \cite{Turyshev-Toth:2017,Turyshev-Toth:2018-grav-shadow,Turyshev-Toth:2019}). Specifically, we required that rays with impact parameters $b\le R_\odot^\star=R_\odot +r_g$ are completely absorbed by the Sun \cite{Turyshev-Toth:2017} and no reflection or coherent reemission occurs. Technically, such formulation relies on the semiclassical relationship between the partial momentum, $\ell$, and the impact parameter, $b$, that is given as as $\ell=k b$ (see relevant discussion in \cite{Turyshev-Toth:2019}.)  Therefore, we require that no outgoing waves (i.e., $\propto H^{+}_\ell$) exist in the region behind the Sun for rays of light with impact parameter $b\leq R_\odot^\star$ or, equivalently, for $\ell \leq kR_\odot^\star$. This results in the Debye potential given by (\ref{eq:Pi-s_a+0}) that is valid for all distances outside the Sun $r>R^\star_\odot$ and all angles.

The expression (\ref{eq:Pi-s_a+0}) for the Debye potential is rather complex. It requires the tools of numerical analysis to fully explore its behavior and the resulting EM field \cite{Kerker-book:1969,vandeHulst-book-1981,Grandy-book-2005}. However, in most practical applications, we only need to know the field in the forward direction. Furthermore, our main interest is to study the largest effect of the solar gravitational field on light propagation, which corresponds to the smallest values of the impact parameter. In this situation, we may simplify the result (\ref{eq:Pi-s_a+0}) by taking into account the asymptotic behavior of the functions $H^{\pm}_\ell(kr_g,kr)$, while considering the resulting EM field at large heliocentric distances, such that $kr\gg\ell$, where $\ell$ is the order of the Coulomb function (see p.~631 of \cite{Morse-Feshbach:1953}). Given the fact that $r_g\ll R^*_\odot$, this approach may be used to describe the EM field immediately outside the solar photosphere all the way to the focal region.

\subsection{Optical properties of the SGL}
\label{sec:general}

The solution for the EM field given by Eqs.~(\ref{eq:DB-sol00p*})--(\ref{eq:Pi-s_a+0}) was analyzed extensively in \cite{Turyshev:2017,Turyshev-Toth:2017,Turyshev-Toth:2019}. The optical properties of the SGL are now well known \cite{Turyshev-Toth:2018,Turyshev-Toth:2018-plasma,Turyshev-Toth:2018-grav-shadow}. Below, we  summarize the most important of these.

\begin{figure}
\includegraphics[scale=0.30]{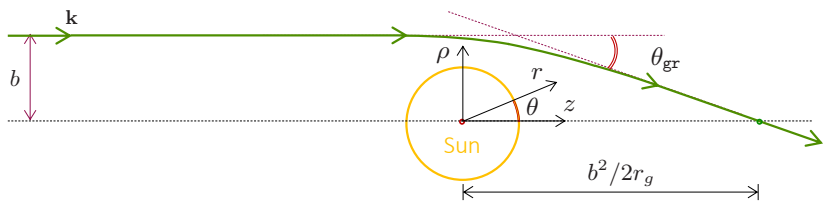}
\caption{\label{fig:go}Focusing of light by the SGL. The heliocentric coordinate system is such that the $z$-axis is along the incoming direction of the wave propagation, given by the wavevector $\vec k$. Spherical ($r,\theta$) and cylindrical ($\rho,z$) coordinates, used in the text, are shown. The azimuthal angle $\phi$ is suppressed. The trajectory of a light ray with impact parameter $b$ with respect to the Sun is deflected towards the Sun by the angle $\theta_{\tt gr}=2r_g/b$, causing it to  intersect the $z$-axis at the heliocentric distance $z=b^2/2r_g$.
}
\end{figure}

\begin{figure}
\includegraphics[scale=0.27]{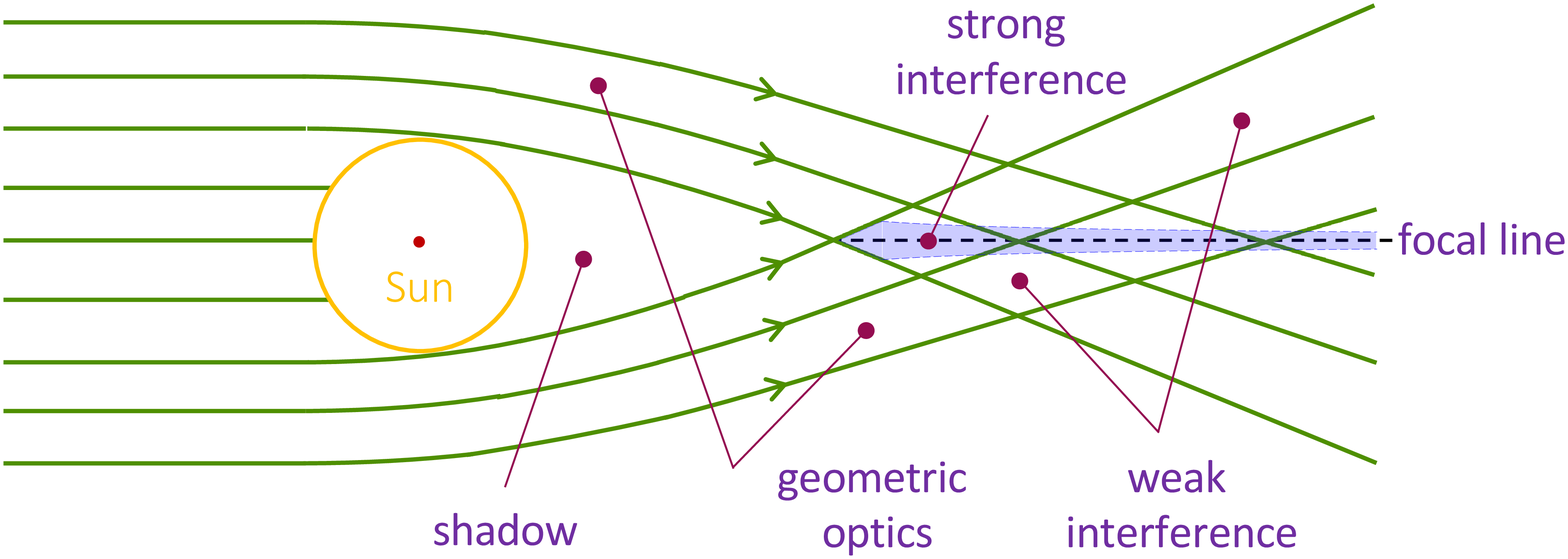}
\caption{\label{fig:regions}The different optical regions of the SGL with respect to light from a source at infinity. Rays with a larger impact parameter intersect at a greater distance, forming a focal half-line (shown by the dashed line). The shaded area is the strong interference region in the immediate proximity of the optical axis as discussed in the text.}
\end{figure}

As was shown in \cite{Turyshev:2017,Turyshev-Toth:2017}, the SGL exists due to the effect of gravitation on the refractive properties of spacetime, focusing light. According to Einstein's general theory of relativity, the trajectory of a photon that travels near the Sun is deflected towards the Sun (its largest effect on light propagation described by the solar gravitational monopole) by the angle of $\theta_{\tt gr}=2r_g/b$, where $r_g=2GM_\odot/c^2$ is the Schwarzschild radius of the Sun and $b$ is the photon's solar impact parameter; see Fig.~\ref{fig:go}. Because solar gravity is weak, the actual deflection angle is very small, so that parallel rays of light passing by the Sun near the solar surface, $b=R_\odot$, focus at the large heliocentric distance of $R_\odot^2/2r_g=547.6~(b/R_\odot)^2$ AU. The SGL does not have a single focal point. Rays with larger impact parameters focus at greater distances from the Sun; thus, a focal half-line forms, as shown in Fig.~\ref{fig:regions}.

\begin{figure}
\includegraphics[scale=0.40]{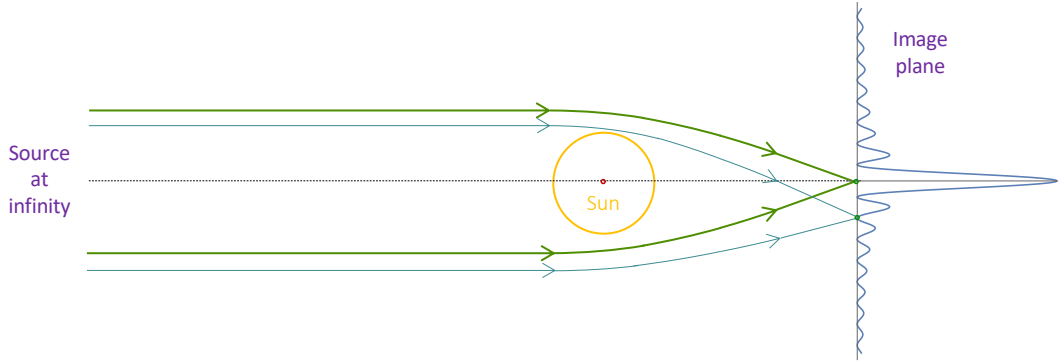}
\caption{\label{fig:geom-wo}Wave-optical treatment of the diffraction of light on the solar gravitation field. In the strong interference region, rays with different impact parameters would have different optical path lengths which leads to formation of the interference pattern (shown conceptually, not to scale) on the image plane. }
\end{figure}

Based on the analysis presented in \cite{Turyshev-Toth:2017, Turyshev-Toth:2019}, we know that diffraction of light on the solar gravitational monopole together with the effect of light's interaction with the Sun results in the formation of four regions behind the Sun  (Fig.~\ref{fig:regions}) with characteristically different EM field behavior, namely:
\begin{enumerate}[i)]
\item Rays with impact parameters $b\leq R^*_\odot$ are completely absorbed by the Sun, resulting in the shadow region directly behind the Sun.  Because of the gravitational bending of light, the shadow has a hyperboloidal shape with its vertex at the heliocentric distance of $R_\odot^2/2r_g=547.6$~AU and the rim touching the Sun. As the edge of the solar disk is not optically smooth, no Arago spot forms along the shadow centerline. Thus, no discernible light from the distant source reaches this region \cite{Turyshev-Toth:2018-grav-shadow}.

\item Most of the area outside the shadow region is characterized by distances $r\geq R^*_\odot$ and the angles  $\theta\gg \sqrt{2r_g/z}$. As the Sun blocks rays with impact parameter $b\leq R^*_\odot$, only one light ray passes through any given point in this region. This is the region of geometric optics. At any given point in this region, the EM field is represented by the incident wave \cite{Turyshev-Toth:2017}. The phase of the incident ray is well described by the geometric optics approximation, thus giving the name for this region. The structure of the EM field here was discussed in \cite{Turyshev-Toth:2017}.

 \item As the light rays propagate towards the focal line, their deflected trajectories define a plane. The path of propagation ultimately begins to intersect trajectories in the same plane, passing by the opposite side of the solar monopole with impact parameter $r\geq R^*_\odot$. The smaller the angle $\theta$ at the intersection point, the smaller the optical path difference (OPD) between the rays. As  $|\theta|\gg \sqrt{2r_g/z}$, from (\ref{eq:mu-point}) we see that the path difference is quite large, namely ${\rm OPD}=\sqrt{{2r_g z}}\,|\theta|\gg 2r_g\gg \lambda$. Although a weak interference pattern forms, the geometric optics approximation remains applicable here. This is what we call the region of the weak interference.  At any given point in this region, the EM field can be represented by the incident and scattered waves (see \cite{Turyshev-Toth:2017} for details). An observer  in this region would see two images of uneven brightness representing the same point source, situated on opposite sides of the Sun. As the point of intersection gets closer to the optical axis, the difference in brightness decreases.

\item For impact parameters $b> R^*_\odot$ and thus the distances beyond 547.6~AU, in the immediate vicinity of the optical axis, $0\leq |\theta| \simeq \sqrt{2r_g/z}$, we enter the region of strong interference, i.e., ``the focal beam of extreme intensity'' \cite{Herlt-Stephani:1976}. Because of azimuthal symmetry, rays with different azimuthal angles (defining different planes of propagation) intersect at or near the focal line. These intersecting rays have the smallest possible ${\rm OPD} \simeq \lambda$, thus creating a strong interference pattern (Fig.~\ref{fig:regions}). The components of the Poynting vector and the EM field intensity are oscillating with a spatial period of $\rho=\sqrt{z/2r_g}\lambda$. If positioned exactly on the optical axis (i.e., $\theta={\rm OPD}=0$), the observer would see a complete Einstein ring with the brightness of the source being greatly amplified. If positioned slightly away from the optical axis (i.e., $0<|\theta|\simeq \sqrt{2r_g/z}$ and, thus,  $\lambda<{\rm OPD}\lesssim r_g$), two incomplete arcs with slightly different angular sizes and intensities (albeit both strongly amplified) would appear. This is the region where the SGL acquires its most impressive optical properties. To describe the image formation processes here, one needs a wave-optical treatment that was developed in \cite{Turyshev:2017,Turyshev-Toth:2017,Turyshev-Toth:2019}.
\end{enumerate}

The strong interference region is of the greatest practical importance. This is the region where the SGL offers major light amplification and high angular resolution, which are both needed for imaging of exoplanets.  Using a wave-optical treatment of light diffraction on the solar gravitational monopole, we established \cite{Turyshev-Toth:2017} that for a point source at infinity and for an observer at the heliocentric distance $z$, the SGL's light amplification factor is given by
{}
\begin{eqnarray}
{\bar \mu}_z&=&
\mu_0 J^2_0\Big(\frac{2\pi}{\lambda}\sqrt{{2r_g z}}\,\theta
\Big),\qquad {\rm with}
\qquad
\mu_0=\frac{4\pi^2}{1-e^{-4\pi^2 r_g/\lambda}}
\frac{r_g}{\lambda}.
\label{eq:mu-point}
\end{eqnarray}

For any given $z$, light amplification reaches its maximum value when $\theta=\rho/z=0$. Away from the optical axis, destructive interference becomes significant as light rays with various optical path lengths converge on the same point. An observer positioned on the optical axis would register a major increase in brightness, characterized by the factor $\sim 4\pi^2 r_g/\lambda$ in (\ref{eq:mu-point}), which, for the wavelength of $\lambda=1~\mu$m, is $\sim 10^{11}$. Elsewhere in the image plane, rays with a non-vanishing difference of their optical paths form the interference pattern, as conceptually shown in Fig.~\ref{fig:geom-wo}. The average amplitude of the intensity of the pattern falls off approximately with the inverse of the distance from the optical axis, characterized by the radial coordinate $\rho\simeq \theta z$.
This level of light amplification is one of the major benefits of the SGL. The SGL's other main advantage is its angular resolution of $\sim$0.5 nanoarcseconds that is determined at the first null of the Bessel function in  (\ref{eq:mu-point}) which occurs  quite close to the optical axis:
$\rho_{\tt SGL0}\simeq  4.5~({\lambda}/{1~\mu{\rm m}})({z}/547.6~{\rm AU})^\frac{1}{2}~{\rm cm}$
\cite{Turyshev-Toth:2017}. These impressive values have motivated us to consider using the SGL for imaging distant, small and faint sources.

As we began the development of the appropriate simulation tools, we realized that  the description above is based on studying the EM field received from a point source that is positioned at the infinite distance from the Sun. However, this assumption is not valid when we attempt to describe the imaging of extended and resolved sources analytically. It became necessary to develop analytical tools exactly for this purpose.  This is discussed next.

\section{EM field from the source at a finite distance}
\label{sec:EM-field-fin-dist}

Our previous work on the wave-optical theory of the SGL relied on developments in atomic physics, dealing with nuclear scattering on a Coulomb potential. The relevant theoretical efforts from the twentieth century provided us with a rich set of mathematical tools and methods \cite{Messiah:1968,Schiff:1968,Morse-Feshbach:1953,Sharma-book:2006,Grandy-book-2005,Friedrich-book-2013} that can be used to study the scattering of light in the solar gravity field. These methods are directly applicable to problems in atomic physics, where the focus is on the asymptotic behavior of a scattered field.
In case of a source at a finite distance, however, these tools require additional development. We can still use the geometric optics approximation to describe light propagation in the region of geometric optics. The solution to the appropriate geodetic equation is well known and describes the trajectory and phase of a light ray along the path from the source at a finite distance to the observer position (see, for instance, discussion in Appendices B.1 and B.2 of \cite{Turyshev-Toth:2017}). However, to describe the optical processes in the weak and, especially, the strong interference regions, this approach is inadequate. We need to solve the time-independent Schr\"odinger equation with a Coulomb potential, as was done in \cite{Turyshev-Toth:2017}.
Technically, the solution must be sought using a form of the incident spherical wave for the source at a finite distance \cite{Born-Wolf:1999} and would involve a set of appropriately modified Coulomb functions.
No exact solution to this problem is known. Instead, we are forced to develop an approximation.

\subsection{The geometry of the problem}
\label{sec:opt-axis}

\begin{figure}
\includegraphics[scale=0.38]{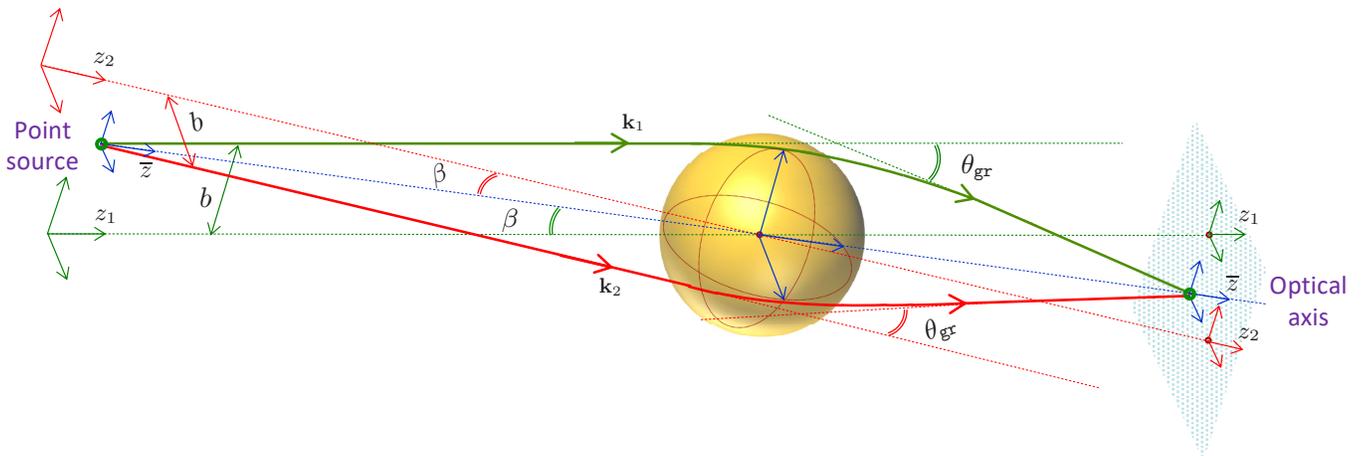}
\caption{\label{fig:point-3D}The three-dimensional geometry of the SGL, focusing light from a point source located at a finite distance. Two rays of light with wavevectors $\vec k_1$ and $\vec k_2$ are shown. The rays move in different planes, which intersect along the optical axis. Note that the $z$-axis is no longer uniquely defined. However, the optical axis $\overline z$ is unique and preserves the axial symmetry.}
\end{figure}

We consider a point source located at a distance $r_0$ from the Sun (Fig.~\ref{fig:point-3D}).  The line connecting the point source and the center of the Sun defines the optical axis of the SGL, $\overline z$. Clearly, there are an infinite number of rays that are emitted by the point source in $4\pi$ steradians; many of these rays travel towards the Sun. Depending on the impact parameter, some these rays will either be absorbed by the Sun or else will travel beyond the Sun, eventually entering the geometric optics and the interference regions. In the spherically symmetric gravitational field (\ref{eq:metric-gen})--(\ref{eq:pot_w_1**}), the optical axis is the axis of axial symmetry. I.e., the geometry of the problem is invariant under a rotation around the $\overline z$-axis.

We introduce a heliocentric spherical coordinate system with the polar axis directed along the  $\overline z$-axis. The coordinates of the point source in this coordinate system are $(r_0,0, 0)$. Next, we take a ray of light emitted by the point source in the direction towards the Sun with the solar impact parameter $b\geq R^*_\odot$. This wavevector $\vec k$ of this ray and the optical $\overline z$-axis define a plane. We set up a $z$-axis in this plane, parallel with $\vec{k}$. This axis corresponds to the $z$-axis that was used to describe the problem with the source being at infinity (shown in Fig.~\ref{fig:go} and discussed in Sec.~\ref{sec:em-waves-gr+pl}). In the spherical coordinate system corresponding to this $z$-axis, the coordinates of the point source are given by $(r_0,b/r_0, \phi_0)+{\cal O}(b^2/r_0^2)$.

At this point, we face a technical challenge. On the one hand, we have the solution (\ref{eq:DB-sol00p*})--(\ref{eq:Pi-s_a+0}), which was obtained under the assumption that the source is at infinity. This solution was obtained in the coordinate system corresponding to the $z$-axis that we just defined. However, to properly describe the problem, we need to find a solution for the ray propagating with the incident wavevector $\vec k$ given in a coordinate system corresponding to the optical $\overline z$-axis associated with a point source at a large, but finite distance. This problem requires one either solving the Maxwell equations in the $\overline z$-coordinate system with an appropriate form of the incident logarithm-modified plane wave (see \cite{Turyshev-Toth:2017} for details) or to transform the solution (\ref{eq:DB-sol00p*})--(\ref{eq:Pi-s_a+0}) from $z$-coordinates to $\overline z$-coordinates. No such solution is currently known.

To develop the needed solution for the EM field, we first note that because of the separation of variables that was used to solve the Maxwell equations (see \cite{Turyshev-Toth:2017} for details),  the solution  (\ref{eq:DB-sol00p*})--(\ref{eq:Pi-s_a+0}) isolates the dependence on the azimuthal angle $\phi$ from the dependence on the other two variables, $r$ and $\theta$, present in $\alpha(r,\theta)$, $\beta(r,\theta)$ and $\gamma(r,\theta)$, given by (\ref{eq:alpha*})--(\ref{eq:gamma*}). Thus, for any given angle $\phi=\phi_0$, the needed solution is already available in the form of (\ref{eq:alpha*})--(\ref{eq:Pi-s_a+0}), but the solution then needs to be rotated by the angle
\begin{equation}
\beta = \frac{b}{r_0}+{\cal O}(b^2/r_0^2),
\label{eq:beta}
\end{equation}
in the plane defined by $\phi=\phi_0$ and containing the optical axis. This rotation transforms the polar angle as
\begin{equation}
{\overline\theta}=\theta+\beta,
\label{eq:otheta}
\end{equation}
but leaves $r\to r+{\cal O}(b^2/r_0^2)$. The azimuthal angle, however, changes according to
\begin{equation}
\tan\overline\phi=\tan\phi+\frac{\sin(\phi-\phi_0)}{\tan\theta\cos^2\phi}\sin\beta+{\cal O}(b^2/r_0^2).
\label{eq:ophi}
\end{equation}

In other words, there is an additional rotation in the azimuthal plane, $\phi\to\overline\phi$. As it turns out, the actual magnitude of this rotation does not need to be computed, for the following reasons:
\begin{enumerate}[1)]
\item In regions other than the region of strong interference, at any point in space, light arrives in the form of at most two rays, both of which travel in the same plane. The intensity of light, therefore, does not depend on $\phi$.
\item In the strong interference region, we are near the optical axis, axial symmetry is restored, and dependence on the transformed azimuthal angle $\overline\phi$ vanishes, which is also apparent from the obvious degeneracy of Eq.~(\ref{eq:ophi}) when $\theta\to 0$. Therefore, although in this region, multiple rays of light traveling in different azimuthal planes are combined, the result remains independent of either $\phi$ or $\overline\phi$.
\end{enumerate}
These considerations allow us to greatly simplify the problem by considering light propagation in a plane only.

Another important concern regarding a point source at a finite distance is that light from that source no longer arrives in the form of a plane wave. However, when the source is at a great distance and we are studying a narrow beam of light, we may use the paraxial approximation, which allows us to continue using the formalism developed for incident plane waves instead of reformulating the problem with spherical waves. Using this approximation, we must rescale the field intensity of the plane wave, $E_0$, to account for its distance dependence from the source, namely $E_0\rightarrow E^{\tt s}_0/r_0$, where $E^{\tt s}_0$ is the field intensity of the corresponding spherical wave.

\subsection{Solving Maxwell's equations}

Considering the solution for the Debye potential given by Eq.~(\ref{eq:Pi-s_a+0}), we observe that it depends on the associated Legendre polynomials, $P^{(1)}_\ell(\cos\theta)$ and the Coulomb--Hankel functions, $H^{\pm}_\ell(kr_g,kr)$. Both of these quantities are given with respect to the $z$-axis, which, as discussed above, is not the most optimal axis to use in the case the source is at a finite distance. Not only is the $z$-axis not uniquely defined but it breaks axial symmetry.

Given the coordinate transformation (\ref{eq:otheta}), we see that the Legendre polynomials written with respect to the $\overline z$-axis may be given as  $P_\ell(\cos\overline \theta)=P_\ell(\cos (\theta+\beta))=P_\ell(\cos \theta)+{\cal O}(\theta b/r_0)$, which is sufficient for our purposes.

\begin{figure}
\includegraphics[scale=0.40]{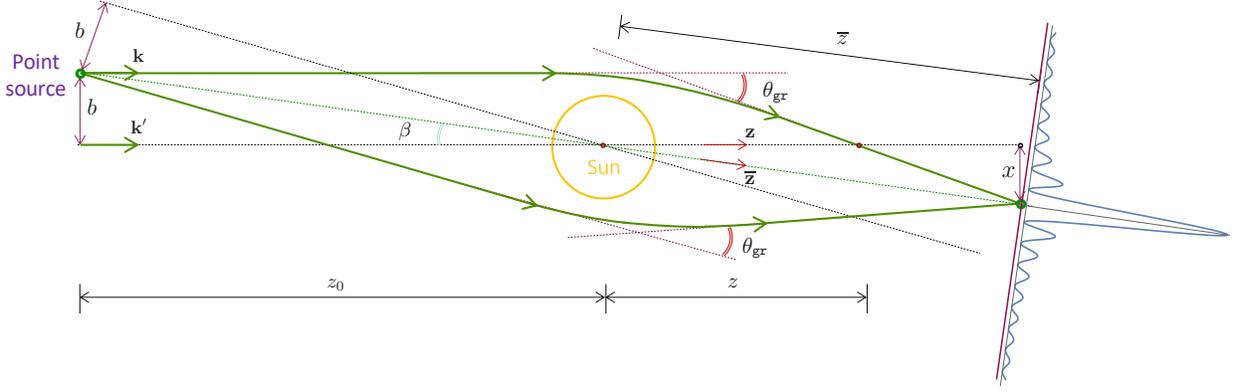}
\caption{\label{fig:geom-opt}The SGL focusing light from a point source located at a finite distance. Two rays of light with the same impact parameter $b$, traveling in the same plane but on opposite sides of the Sun,
are shown. The incident rays are no longer parallel. The diagram is arranged such that the top incident ray appears horizontal, as in Fig.~\ref{fig:go}. The wavevector $\vec k$ is inclined with respect to the optical axis $\overline z$ by the angle $\beta=b/z_0$. Both rays intersect $\overline z$ at the distance $\overline z=(b^2/2r_g)(1+b^2/2r_g z_0)$.
}
\end{figure}

Next,  we turn our attention to the solution for the radial function in (\ref{eq:Pi-s_a+0})  which is given by the Coulomb--Hankel functions $H^{\pm}_\ell(kr_g,kr)$. The asymptotic form of these functions has to account for the fact that the light source is at a finite distance, at spherical coordinates $(r_0, b/r_0, \phi_0)$. We follow the path of propagation of a light ray in the plane $\phi=\phi_0$. The corresponding asymptotic behavior of $H^{\pm}_\ell(kr_g,kr)$ was established in Appendix \ref{sec:rad_eq_wkb} for  $kr\rightarrow\infty $ and $r\gg r_{\tt t}=\sqrt{\ell(\ell+1)}/k$ (see \cite{Turyshev-Toth:2017,Turyshev-Toth:2018-plasma}) in the form of (\ref{eq:R_solWKB+=_bar-imp}) and is given by
{}
\begin{eqnarray}
\lim_{kr\rightarrow\infty} H^{\pm}_\ell(kr_g,kr)&\sim&
\exp\Big[\pm ik\big(r+r_0+r_g\ln(4k^2rr_0)\big)+\frac{\ell(\ell+1)}{2k}\Big(\frac{1}{r}+\frac{1}{r_0}\Big)+\sigma_\ell-\frac{\pi \ell}{2}\Big)\Big] +{\cal O}\big((kr)^{-2}, r_g^2\big),
\label{eq:Fass*}
\end{eqnarray}
which includes the contribution from the centrifugal term, $\propto \ell(\ell+1)/2kr$, in the radial equation for the Debye potential (see Appendix \ref{sec:rad_eq_wkb} here or Appendix A in \cite{Turyshev-Toth:2018}). Including the centrifugal term lets us better describe the bending of the light ray's trajectory under the influence of solar gravity. In addition, Eq.~(\ref{eq:Fass*}) contains terms that describe the dependence of the total phase of the EM wave on the distance to the source, $r_0$.

Furthermore, as the optical axis preserves the axial symmetry, all the rays emitted towards the Sun from a point on that axis with coordinates $(r_0,0, \phi_0)$ and impact parameter  $b>R^*_\odot$, will intersect the optical axis behind the Sun at one point, at the heliocentric distance of  $\overline z=b^2/2r_g(1+b^2/2r_g r_0)$. As a result of this axial symmetry, the solution does not depend on the specific choice of $\phi_0$.

The presence of the $\propto \ell^2$ term in (\ref{eq:Fass*}) (and also in (\ref{eq:R_solWKB+=_bar-imp})) is important is it allows for a better description of the  light ray's trajectory. Taking the semiclassical representation of the partial momenta via the impact parameter as $\ell=kb$, for $\ell\gg1$ we may present  the Euclidean part of the phase of (\ref{eq:Fass*}) as
\begin{equation}
k(r+r_0)+\frac{\ell(\ell+1)}{2k}\Big(\frac{1}{r}+\frac{1}{r_0}\Big)=
k\big(r+\frac{b^2}{2r}\big)+k\big(r_0+\frac{b^2}{2r_0}\big)
= k\sqrt{r^2+b^2}+k\sqrt{r^2_0+b^2}+{\cal O}(b^4/r^4,b^4/r_0^4),
\end{equation}
which now correctly describes  the Euclidean distance that the light travels from its emission at $r_0$ to the point of its detection, $\overline z$.  The remaining part of the phase in (\ref{eq:Fass*}) that is given as $kr_g\ln(4k^2rr_0)$ is due to the lengthening of the light ray's path in the curved spacetime that is induced by the solar gravitational field.

Finally, as in the realistic exoplanet imaging situations, the ratio $\beta=b/r_0$ is small, any effect on the amplitude of the EM field from the relevant $\beta$-rotation of the EM field is negligible. As we shall see in Sec.~\ref{sec:go-em-outside}, most of the contribution from $\beta$ affects the phase of the EM wave, which is  our primary interest.

\subsection{EM field in the shadow region}
\label{sec:shadow}

In the shadow behind the Sun (i.e., for impact parameters $b\leq R^\star_\odot$, see Fig.~\ref{fig:regions}), the EM field is represented by the Debye potential of the shadow, $\Pi_{\tt sh}$, which, from (\ref{eq:Pi-s_a+0}),  is determined solely by the incoming wave, as prescribed by the fully absorbing boundary conditions:
{}
\begin{eqnarray}
\Pi_{\tt sh} (r, \theta)&=&-
\frac{E_0}{2ik^2}\frac{u}{r}\sum_{\ell=1}^\infty i^{\ell-1}\frac{2\ell+1}{\ell(\ell+1)}e^{i\sigma_\ell}H^-_\ell(kr_g, kr)P^{(1)}_\ell(\cos\theta).
  \label{eq:Pi_sh}
\end{eqnarray}

As discussed in \cite{Turyshev-Toth:2017,Turyshev-Toth:2018-grav-shadow}, the potential (\ref{eq:Pi_sh}), to the required level of accuracy, produces no EM field in the area $-\pi/2\le \theta\le\pi/2$. In other words, there is no light in the shadow. Additionally, one can show that even outside the shadow region behind the Sun, the potential (\ref{eq:Pi_sh}) results in a very small EM field, negligible for our analysis \cite{Turyshev-Toth:2018,Turyshev-Toth:2018-grav-shadow}. As a result, we omit this term from the Debye potential (\ref{eq:Pi-s_a+0}) in our analysis when we discuss the EM field in the geometric optics and interference regions. Therefore, we focus on the EM field produced only by the outgoing part of the Debye potential (\ref{eq:Pi-s_a+0}), namely $\propto H^+_\ell(kr_g, kr)$.

\subsection{EM field outside the shadow}
\label{sec:EM-field}

In the region outside the solar shadow (i.e., for light rays with impact parameters $b>R_\odot^\star$), which includes  the geometric optics  and both interference regions (see Fig.~\ref{fig:regions}), the EM field is derived from the Debye potential (\ref{eq:Pi-s_a+0}). For this, we substitute (\ref{eq:Fass*}) in (\ref{eq:Pi-s_a+0}) and derive the Debye potential without the incident wave (discussed in Sec.~\ref{sec:shadow}):
{}
\begin{eqnarray}
\Pi(r, \theta)&=& -\frac{E_0}{k^2}\frac{u}{r}e^{ik(r+r_0+r_g\ln 4k^2rr_0)}\sum_{\ell=kR^\star_\odot}^\infty \ell^{-1}
e^{i\big(2\sigma_\ell+\frac{\ell^2}{2k}(\frac{1}{r}+\frac{1}{r_0})\big)}
P^{(1)}_\ell(\cos\theta) +{\cal O}(r_g^2, (kr)^{-3}),
\label{eq:Pi-0+}
\end{eqnarray}
where we recognize that for large $\ell\geq kR^\star_\odot$, we may replace $\ell+1\rightarrow \ell$ and $\ell+\textstyle\frac{1}{2}\rightarrow \ell$. Expression (\ref{eq:Pi-0+}) is the Debye potential that yields the EM field in the regions of geometric optics and the interference region.

Introducing, for convenience, the effective distance $\tilde r$ in the form
\begin{equation}
\frac{1}{\tilde r}=\frac{1}{r}+\frac{1}{r_0} \qquad \Rightarrow \qquad
{\tilde r}=\frac{rr_0}{r+r_0},
 \label{eq:r-tilde}
 \end{equation}
we derive the components of the EM field. For this, we  use (\ref{eq:Pi-0+}) and (\ref{eq:r-tilde}) in the expressions (\ref{eq:alpha*})--(\ref{eq:gamma*}) to derive the factors $\alpha(r,\theta), \beta(r,\theta)$ and $\gamma(\theta)$, which, to the order of ${\cal O}\big(r^2_g,(kr)^{-3}\big)$, are computed as
{}
\begin{eqnarray}
\alpha(r,\theta) &=& -E_0\frac{ue^{ik(r+r_0+r_g\ln 4k^2rr_0)}}{k^2r^2}\sum_{\ell=kR^\star_\odot}^\infty \ell e^{i\big(2\sigma_\ell+\frac{\ell^2}{2k\tilde r}\big)}
\Big\{1-\frac{\ell^2}{4u^2k^2r^2}
\Big\}P^{(1)}_\ell(\cos\theta),
  \label{eq:alpha*1*}\\
\beta(r,\theta) &=& E_0\frac{ue^{ik(r+r_0+r_g\ln 4k^2rr_0)}}{ikr}\sum_{\ell=kR^\star_\odot}^\infty\ell^{-1}e^{i\big(2\sigma_\ell+\frac{\ell^2}{2k\tilde r}\big)}
\Big\{\frac{\partial P^{(1)}_\ell(\cos\theta)}
{\partial \theta}\Big(1-\frac{\ell^2}{2u^2k^2r^2}
\Big)+\frac{P^{(1)}_\ell(\cos\theta)}{\sin\theta}
 \Big\},
  \label{eq:beta*1*}\\
\gamma(r,\theta) &=& E_0\frac{ue^{ik(r+r_0+r_g\ln 4k^2rr_0)}}{ikr}\sum_{\ell=kR^\star_\odot}^\infty
\ell^{-1}e^{i\big(2\sigma_\ell+\frac{\ell^2}{2k\tilde r}\big)}
\Big\{\frac{\partial P^{(1)}_\ell(\cos\theta)}
{\partial \theta}+\frac{P^{(1)}_\ell(\cos\theta)}{\sin\theta}\Big(1-
\frac{\ell^2}{2u^2k^2r^2}\Big) \Big\}.~~~~
  \label{eq:gamma*1*}
\end{eqnarray}
Here we neglected small terms that behave as $\propto i/(u^2kr)$; terms $\propto ikr_g/\ell^2$ were also omitted because of the large partial momenta involved, $\ell\geq kR^\star_\odot$. Terms in both of these groups are negligably small when compared to the leading terms in each of these expressions above (a similar conclusion was reached in \cite{Turyshev-Toth:2018-plasma,Turyshev-Toth:2019}.)

Expressions (\ref{eq:alpha*1*})--(\ref{eq:gamma*1*})  represent an important result, allowing us to describe the EM field in the regions of interest for the SGL, namely the geometric optics region and the interference region.

\section{EM field in the  geometric optics and weak interference regions}
\label{sec:go-em-outside}

We are interested in the area that can be reached by light rays with impact parameters $b>R^\star_\odot$ and located behind the Sun at heliocentric distances $r> R^\star_\odot$. This is the region of geometric optics for which
 angles $\theta$ are rather large, satisfying the condition $|\theta|\gg \sqrt{2r_g/r}$ \cite{Turyshev-Toth:2017}. To establish the EM field (\ref{eq:DB-sol00p*}) in this region, we need to develop expressions (\ref{eq:alpha*1*})--(\ref{eq:gamma*1*}) evaluating them to the appropriate level of accuracy. We begin with the investigation of $\alpha(r,\theta)$ from (\ref{eq:alpha*1*}).

\subsection{Solution for the function $\alpha(r,\theta)$ and the radial components of the EM field}
\label{sec:radial-comp}

To evaluate the expression  for $\alpha(r,\theta)$ in the region of geometric optics and, thus, for $\theta \gg \sqrt{2r_g/r}$, we use the asymptotic representation for $P^{(1)}_l(\cos\theta)$ \cite{Bateman-Erdelyi:1953,Korn-Korn:1968,Kerker-book:1969}, valid when $\ell\to\infty$:
{}
\begin{align}
P^{(1)}_\ell(\cos\theta)  &=
\dfrac{-\ell}{\sqrt{2\pi \ell \sin\theta}}\Big(e^{i(\ell+\frac{1}{2})\theta+i\frac{\pi}{4}}+e^{-i(\ell+\frac{1}{2})\theta-i\frac{\pi}{4}}\Big)+{\cal O}(\ell^{-\textstyle\frac{3}{2}}), ~~~~~\textrm{for}~~~~~ 0<\theta<\pi.
\label{eq:P1l<}
\end{align}

With the approximation above, we may replace the sum in the (\ref{eq:alpha*1*})
with an integral yielding an expression for $\alpha(r, \theta)$ that to the order of $
{\cal O}\big(r^2_g,(kr)^{-3}\big)$ has the form:
{}
\begin{eqnarray}
\alpha(r, \theta)&=&\frac{E_0u}{k^2r^2}e^{ik(r+r_0+r_g\ln4k^2rr_0)}\hskip-4pt
\int_{\ell=kR_\odot^\star}^\infty \hskip 0pt
\frac{\ell\sqrt{\ell}d\ell}{\sqrt{2\pi \sin\theta}}\Big(1-\frac{\ell^2}{4u^2k^2r^2} \Big)
e^{i\big(2\sigma_\ell+\frac{\ell^2}{2k\tilde r}\big)}
\Big(e^{i(\ell\theta+\frac{\pi}{4})}+e^{-i(\ell\theta+\frac{\pi}{4})}\Big).
\label{eq:Pi_s_exp1}
\end{eqnarray}
We note that integrating over $\ell$ from $kR_\odot^\star$ to infinity is equivalent to integrating over the impact parameter $b$ ranging from grazing the Sun at $b=R_\odot^\star$ to infinity. We evaluate this integral by the method of stationary phase \cite{Turyshev-Toth:2017}, which is suitable for the evaluation of oscillatory integrals of the type
{}
\begin{equation}
I=\int A(\ell)e^{i\varphi(\ell)}d\ell, \qquad
\ell\in\mathbb{R},
\label{eq:stp-1}
\end{equation}
where the amplitude $A(\ell)$ is a slowly varying function of $\ell$, while $\varphi(\ell)$ is a rapidly varying function of $\ell$.
The integral (\ref{eq:stp-1}) may be replaced, to good approximation, with a sum over the points of stationary phase, $\ell_0\in\{\ell_{1,2,..}\}$, for which $d\varphi/d\ell=0$. Defining $\varphi''=d^2\varphi/d\ell^2$, we obtain the integral
{}
\begin{equation}
I\simeq\sum_{\ell_0\in\{\ell_{1,2,..}\}} A(\ell_0)\sqrt{\frac{2\pi}{\varphi''(\ell_0)}}e^{i\big(\varphi(\ell_0)+{\textstyle\frac{\pi}{4}}\big)}.
\label{eq:stp-2}
\end{equation}

Following  \cite{Turyshev-Toth:2018-plasma}, we see that the relevant $\ell$-dependent part of the phase in (\ref{eq:Pi_s_exp1})  is of the form
{}
\begin{equation}
\varphi_{\pm}(\ell)=\pm\big(\ell\theta+\textstyle{\frac{\pi}{4}}\big)+2\sigma_\ell +\dfrac{\ell^2}{2k \tilde r}+{\cal O}\big(r_g^2, (kr)^{-3}\big).
\label{eq:S-l}
\end{equation}

Similarly to the approach used in \cite{Turyshev-Toth:2018-grav-shadow,Turyshev-Toth:2019}, we evaluate $\sigma_\ell$  for $\ell\gg kr_g$ as:
{}
\begin{eqnarray}
\sigma_\ell&=& -kr_g\ln \ell.
\label{eq:sig-l*}
\end{eqnarray}

The phase is stationary when $d\varphi^{[0]}_{\pm}/d\ell=0$, which, together with (\ref{eq:sig-l*}), implies
{}
\begin{equation}
\pm\theta-\frac{2kr_g}{\ell}+\frac{\ell}{k\tilde r}={\cal O}\big(r_g^2, (kr)^{-3}\big).
\label{eq:S-l-pri=}
\end{equation}

Relying on the semiclassical approximation that connects the partial momentum, $\ell$, to the impact parameter, $b$,
{}
\begin{equation}
\ell\simeq kb,
\label{eq:S-l-pri-p-g}
\end{equation}
for small angles $\theta$ (or, large heliocentric distances, $R_\odot/r<b/r\ll 1$),  we see that the points of stationary phase that must satisfy (\ref{eq:S-l-pri=})  are given by (see \cite{Turyshev-Toth:2017} for details):
{}
\begin{equation}
\frac{b}{r}  +\frac{b}{r_0} = \mp{ {\theta}}+\frac{2r_g}{b}+{\cal O}(\theta^3, r_g^2),
\label{eq:S-l-pri*}
\end{equation}
which describes hyperbolae representing the geodesic trajectories of light rays in the post-Newtonian gravitational field of a mass monopole \cite{Turyshev-Toth:2017}. For impact parameters $b\geq R_\odot^\star$, these trajectories are outside the Sun, crossing from the geometric optics region behind the Sun into the interference region. In essence, (\ref{eq:S-l-pri*})  is a classical thin lens equation that is familiar from geometric optics \cite{Sharma-book:2006,Nambu:2013b,Turyshev-Toth:2017}.  Similarly to the approach demonstrated in \cite{Turyshev-Toth:2019}, the validity of this expression may be extended to higher powers of the small angle $\theta$, yielding complete trigonometric identities.

The result given by (\ref{eq:S-l-pri*}) differs from a similar expression given in \cite{Turyshev-Toth:2017,Turyshev-Toth:2019} for a source located at infinity. The finite distance to the source is captured by the term $\beta=b/r_0$, defined in Eq.~(\ref{eq:beta}); see also Fig.~\ref{fig:geom-opt}. In the coordinate system rotated by angle $\beta$ in the $\phi=\phi_0$ plane, we can easily see that light rays intersect the optical axis at slightly greater heliocentric distances.
For any given impact parameter, the focal point of the SGL, located at $r=b^2/2r_g$ for a source at infinity ($\beta=0$), is shifted $\overline r=b^2/(2r_g)(1+b^2/(2r_g z_0))$. The extra distance that a light ray needs to propagate before it intersects the optical axis is $\delta r=(b^2/2r_g)^2/z_0$, which, for nominal values of the parameters, is computed to be $\delta r=0.05 \,(b/R_\odot)^4(30\, {\rm pc}/z_0)$~AU. The extra distance is small but nonvanishing. The heliocentric distance to the focal point associated with a source at a finite distance, calculated as $\overline r$ with respect to the optical axis, is related to the heliocentric distance $r$ of the focal point associated with a source at infinity by
{}
\begin{equation}
\overline r=r(1+r/r_0)+{\cal O}(r^3/r^2_0).
\label{eq:r-bar}
\end{equation}
 Equation~(\ref{eq:r-bar}) represents a rescaling of all relevant results by an extra a factor that depends on the distance to the source. This mapping between the distances provides an interpretation of the results that we obtain below.

\subsection{Incident and scattered waves}

We now continue to investigate (\ref{eq:S-l-pri*}).  For small but finite angles, $|\theta|\gg \sqrt{2r_g/r}>0$, and large impact parameters, $b\geq R^\star_\odot\gg r_g$, Eq.~(\ref{eq:S-l-pri*}) yields two families of solutions for the points of stationary phase:
{}
\begin{equation}
b_{\tt in}= \mp  \Big(\tilde r \theta+\frac{2r_g}{\theta}\Big)+{\cal O}(\theta^3,r_g^2), \qquad {\rm and} \qquad
b_{\tt s}= \pm \frac{2r_g}{\theta}+{\cal O}(\theta^3,r_g^2),
\label{eq:S-l-pri}
\end{equation}
where the family $b_{\tt in}$ represents the incident wave with light ray trajectories bent towards the Sun, obeying the eikonal approximation of geometric optics and the family $b_{\tt s}$ describes the scattered wave.

\begin{figure}
\includegraphics[scale=0.40]{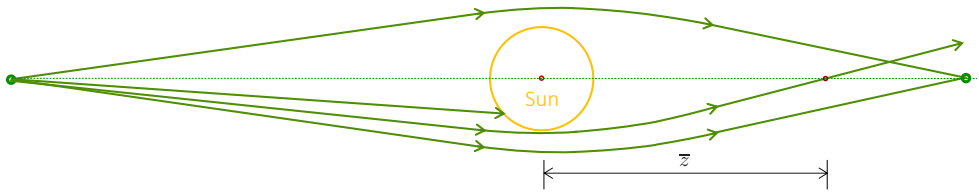}
\caption{\label{fig:regions-fin}The incident and scattered rays in the case of finite distance to the source. Scattered rays with impact parameter $|b_{\tt s}|<R^*_\odot$ (shown by the third line from the bottom) are blocked by the Sun, resulting in the formation of the shadow and the geometric optics regions (as shown in Fig.~\ref{fig:regions}). Scattered rays with $|b_{\tt s}|>R^*_\odot$ (first and second lines from the bottom) pass by the Sun and, after crossing the optical axis at $\overline z$ or beyond, intersect incident rays, leading to the formation of the weak interference region. When the intersection occurs near or on the optical axis, the incident and scattered rays have similar optical path lengths, leading to the formation of the strong interference region.
}
\end{figure}

The solution (\ref{eq:S-l-pri}) offers a very nice representation of the gravitational scattering of light.  Consider the first term in $b_{\tt in}$, given as $b_0= \tilde r \theta\geq R^*_\odot$. This term describes the propagation of light at the distance $b_0$ from the optical axis that is representative for an empty (or Euclidian) space-time. The second term in $b_{\tt in}$, is identical to $b_{\tt s}$ in magnitude, but has the opposite sign. This second term describes the  scattering of light in the presence of the solar gravitational monopole. For a given impact parameter with respect to the optical axis, $b_0$, as the heliocentric distance $r$ increases, the magnitude of the angle $\theta$ decreases. As $\theta$ gets smaller,  the scattering term becomes more significant, effectively deflecting the trajectory of the light ray. At this point we are still in the geometric optics region (see Figs.~\ref{fig:regions} and \ref{fig:regions-fin}) characterized by only one ray of light from the point source passing through any given point of space. For $\theta\geq 2r_g/R^*_\odot$, the impact parameter for the scattered wave is $|b_{\tt s}|\leq R^*_\odot$ and therefore, the scattered wave is blocked by the Sun, as prescribed by the fully absorbing boundary conditions \cite{Turyshev-Toth:2017}.

Closer to the optical axis, where $\theta<2r_g/R^*_\odot$, the impact parameter of the scattered ray is large enough: $|b_{\tt s}|=2r_g/\theta>R^*_\odot$, so that the scattered ray is no longer blocked by the Sun. As the solution for $b_{\tt s}$ has the sign opposite to that of $b_{\tt in}$, the scattered ray enters the region behind the Sun on the opposite side. After it crosses the optical axis, the scattered ray intersects the incident ray. As the two rays have a drastically different optical paths, no notable interference pattern emerges. An observer would see these two rays correspond to two images of the same source with uneven brightness on opposite sides of the Sun. This is characteristic of the region of weak interference, shown in  Figs.~\ref{fig:regions} and \ref{fig:regions-fin}.

As a result, the two families of solutions (\ref{eq:S-l-pri}) represent the incident and scattered part of the same EM wave shown by Eq.~(21) of \cite{Turyshev-Toth:2017}. This provides support for the interpretation given in \cite{Deguchi-Watson:1987}, where these two solutions were interpreted as distinct rays of light on opposite sides of the Sun. (Note that the strong interference region is not covered by the approximation (\ref{eq:P1l<}). It is described in detail in Sec.~\ref{sec:IF-region}.) As discussed in \cite{Turyshev-Toth:2017,Turyshev-Toth:2018-grav-shadow,Turyshev-Toth:2019}, the presence of both of these families of light rays determines the physical properties of the EM field in three the regions relevant for the SGL, namely the shadow, the geometric optics region and both  interference regions. In addition, the `$\pm$' or `$\mp$' signs represent light rays that propagate on opposite sides of the Sun, as a manifestation of the existing axial symmetry.

\subsection{Computing $\alpha_{\tt in}(r,\theta)$ and $\alpha_{\tt s}(r,\theta)$}

By extending the asymptotic expansion of $H^{+}_\ell(kr_g,kr)$ from (\ref{eq:Fass*}) to the order of ${\cal O}((kr)^{-(2n+1)})$ (i.e., using the WKB approximation as was done in Appendix \ref{sec:rad_eq_wkb}), the validity of the result (\ref{eq:S-l-pri}) may be extended to ${\cal O}(\theta^{2n+1})$. This fact was observed in \cite{Turyshev-Toth:2018-plasma} and used to improve the solution by including terms of higher order in $\theta$.

The first family of solutions of (\ref{eq:S-l-pri}), yielding $\ell_{\tt in}=kb_{\tt in}$,  allows us to compute the phase for the points of stationary phase (\ref{eq:S-l}) for the EM waves moving towards the interference region (a similar calculation was done in \cite{Turyshev-Toth:2018-plasma}):
{}
\begin{eqnarray}
\varphi_{\pm}(\ell_{\tt in})&=& \pm\textstyle{\frac{\pi}{4}}-\textstyle{\frac{1}{2}}k\tilde r\theta^2-kr_g\ln k^2\tilde r^2\theta^2+{\cal O}(kr\theta^4,kr_g\theta^2).
\label{eq:S-l2p}
\end{eqnarray}
From (\ref{eq:S-l}) we  compute $\varphi''(\ell)$ as
\begin{eqnarray}
\dfrac{d^2\varphi_{\pm}}{d\ell^2} &=& \dfrac{1}{k\tilde r}+\frac{2kr_g}{\ell^2} +{\cal O}\big((kr)^{-3}\big).
\label{eq:S-l2+}
\end{eqnarray}
After substituting $\ell_{\tt in}$ in (\ref{eq:S-l2+}), we derive $\varphi''(\ell_{\tt in})= {d^2\varphi_{\pm}}/{d\ell^2} \big|_{\ell=\ell_{\tt in}} $ which yields
\begin{eqnarray}
\sqrt{\frac{2\pi}{\varphi''(\ell_{\tt in})}}&=&\sqrt{2\pi k\tilde r}\Big\{1-\frac{r_g}{\tilde r\theta^2}
+{\cal O}(\theta^2,\frac{r_g}{r}\theta^2)\Big\}.
\label{eq:sf*}
\end{eqnarray}

Now, using (\ref{eq:sf*}), we have the amplitude of the integrand in (\ref{eq:Pi_s_exp1}), for $\ell_{\tt in}\gg1$, taking the form
{}
\begin{eqnarray}
\frac{\ell_{\tt in}\sqrt{\ell_{\tt in}}}{\sqrt{2\pi \sin\theta}}\Big(1-\frac{\ell^2_{\tt in}}{4u^2k^2r^2} \Big)
\sqrt{\frac{2\pi}{\varphi''(\ell_{\tt in})}}
&=&
(\mp1)^{\textstyle\frac{3}{2}} k^2 r\tilde ru^{-2}\sin(\theta+\beta)\Big(1+\frac{r_g}{r(1-\cos(\theta+\beta))}+{\cal O}(\theta^3, \beta^3,\frac{r_g}{r}\theta^2)\Big),\label{eq:S-l3p+*}
\end{eqnarray}
where the angle $\beta$ is from (\ref{eq:beta}) and we used the definition (\ref{eq:r-tilde}) for $\tilde r$  and the solution (\ref{eq:S-l-pri}) for the impact parameter $b_{\tt in}$, namely
{}
\begin{eqnarray}
\tilde r\theta&=&r\big(\theta-\frac{r\theta}{r_0}\big)+{\cal O}(r^3/r_0^2)=r\big(\theta+\frac{b}{r_0}\big)+{\cal O}(r^3/r_0^2)=r\big(\theta+\beta\big)+{\cal O}(r^3/r_0^2).
\label{eq:tildebeta}
\end{eqnarray}

As a result, the expression for $\alpha_{\tt in}(r,\theta)$ from (\ref{eq:Pi_s_exp1}) for the incident wave  takes the form
{}
\begin{eqnarray}
\alpha_{\tt in}(r,\theta)&=&
E_0u^{-1}\frac{r_0}{(r+r_0)}\sin(\theta+\beta) \Big(1+\frac{r_g}{r\big(1-\cos(\theta+\beta)\big)}+{\cal O}(\theta^2,\frac{r_g}{r}\theta^2)\Big)e^{i \varphi_{\tt in}(r,\theta)},
\label{eq:Pi_s_exp4+1*}
\end{eqnarray}
where to derive the phase, $\varphi_{\tt in}(r,\theta)$, according to (\ref{eq:stp-2}), we combined the WKB phase from the exponent in front of the integral in (\ref{eq:Pi_s_exp1}) and the $\ell$-dependent contribution from (\ref{eq:S-l2p}):
$k(r+r_0+r_g\ln4k^2rr_0)+ \varphi_{\pm}(\ell_{\tt in})+\textstyle{\frac{\pi}{4}}
$,  yielding
{}
\begin{eqnarray}
\varphi_{\tt in}(r,\theta)&=&
k\Big(r_0\cos\beta+r_g\ln 2kr_0+r\cos(\theta+\beta)-r_g\ln kr\big(1-\cos(\theta+\beta)\big)\Big).
\label{eq:phase_rr0}
\end{eqnarray}
We define $\vec k_\beta=-{\vec n}_0$ being the vector along the direction from the point source to the center of the Sun, i.e., along the optical axis. Then, the following identity holds $2kr_0= kr_0-k(\vec k_\beta\cdot\vec r_0)$. With this, the expression (\ref{eq:phase_rr0}) takes a familiar form:
{}
\begin{eqnarray}
\varphi_{\tt in}(r,\theta)&=&k\Big(
\vec k_\beta \cdot(\vec x-\vec x_0)-r_g\ln \frac{r-({\vec k}_\beta\cdot\vec x)}{r_0-(\vec k_\beta\cdot\vec x_0)}\Big),
\label{eq:phase_rr0-vec}
\end{eqnarray}
that describes the phase of an EM wave obtained with the geodesic equation (discussed in Appendix B of Ref.~\cite{Turyshev-Toth:2017}, with relevant results given there by (B22) and (B33), correspondingly.)  The results (\ref{eq:Pi_s_exp4+1*})--(\ref{eq:phase_rr0}) represent good evidence that in the case of a finite distance to the source, the overall solution for the EM field is rotated by the angle $\beta$, aligning it with the optical axis specified by a point source with coordinates $({\vec b},r_0)$ and the center of the Sun. However, an obvious difference compared to a solution of the geodesic equation is the fact that (\ref{eq:phase_rr0-vec}) describes the evolution of the phase along the trajectory with a zero impact parameter.

Result (\ref{eq:phase_rr0-vec}) justifies our approach of modifying the existing solution for the EM field by applying the eikonal approximation, as discussed in Sec.~\ref{sec:opt-axis}.   The use of the optical axis $\overline z$ restores the axial symmetry of the problem and the entire EM field representing the family of rays emitted towards the Sun with the same impact parameter $b$, may now be obtained by a simple rotation around the $\overline z$-axis by the angle $\phi_0$.

Now we consider the second family of solutions in (\ref{eq:S-l-pri}), given by $\ell_{\tt s}=kb_{\tt s}$ (similar derivations were made in \cite{Turyshev-Toth:2018-grav-shadow}), which allow us to compute the stationary phase as
{}
\begin{eqnarray}
\varphi_{\pm}(\ell_{\tt s})&=&
\pm\textstyle{\frac{\pi}{4}}-kr_g\ln 2k\tilde r+kr_g\ln k\tilde r \textstyle{\frac{1}{2}}\theta^2-2kr_g\ln \frac{kr_g}{e}+{\cal O}(kr_g \theta^2).
\label{eq:S-l27*+}
\end{eqnarray}

Using (\ref{eq:S-l2+}) and $\ell_{\tt s}=kb_{\tt s}$ from (\ref{eq:S-l-pri}), we compute the second derivative of the phase with respect to $\ell$:
{}
\begin{equation}
\varphi''_{\pm}(\ell_{\tt s})=\frac{\theta^2}{2kr_g}\Big(1+\frac{2r_g}{\tilde r\theta^2}\Big)+{\cal O}(\theta^3), \qquad {\rm thus, } \qquad
\sqrt{\frac{2\pi}{\varphi''(\ell_{\tt s})}}=\frac{\sqrt{4\pi kr_g}}{\theta}\Big(1-\frac{r_g}{\tilde r\theta^2}\Big).
\label{eq:S-l2202}
\end{equation}

At this point, we may evaluate the amplitude of the integrand in (\ref{eq:Pi_s_exp1}), which, for $\ell_{\tt s}\gg1$,  is given as
{}
\begin{eqnarray}
\frac{\ell_{\tt s}\sqrt{\ell_{\tt s}}}{\sqrt{2\pi \sin\theta}} \Big(1-\frac{\ell^2_{\tt s}}{4u^2k^2r^2} \Big)\sqrt{\frac{2\pi}{\varphi''(\ell_{\tt s})}}&=&
(\mp1)^{\textstyle\frac{3}{2}}\Big( \frac{2kr_g}{\theta}\Big)^2\frac{1}{\theta}\Big(1-\frac{r_g}{\tilde r\theta^2}\Big).
\label{eq:S-l3p+*=}
\end{eqnarray}

As a result, the expression for $\alpha_{\tt s}(r,\theta)$ representing the scattered wave in (\ref{eq:Pi_s_exp1})  takes the form
{}
\begin{eqnarray}
\alpha_{\tt s}(r,\theta)&=&-E_0\Big(\frac{2r_g}{r}\Big)^2\frac{1}{\theta^3}
e^{ik\big(r_0+r_g\ln 2kr_0+r+r_g\ln kr(1-\cos(\theta+\beta))\big)+2i\sigma_0}
\sim {\cal O}(r_g^2),
\label{eq:alpha-2}
\end{eqnarray}
where the phase was computed by combining the WKB phase from (\ref{eq:Pi_s_exp1}) and the $\ell$-dependent contribution (\ref{eq:S-l27*+}), as we did to compute (\ref{eq:phase_rr0}).  We conclude that to the order of ${\cal O}(r_g^2)$, there is no scattered wave in the radial direction, which is consistent with the results reported in \cite{Turyshev-Toth:2017}, extending those to include dependence on the source position.

The results (\ref{eq:Pi_s_exp4+1*}), (\ref{eq:alpha-2}) are the radial components of the EM wave corresponding to the two families of the impact parameters (\ref{eq:S-l-pri}). We use these solutions to discuss the EM field in the geometric optics region.

\subsection{Evaluating the  function $\beta(r,\theta)$}
\label{sec:amp_func-beta}

To evaluate the magnitude of the function $\beta(r, \theta)$ in (\ref{eq:beta*1*}), we  rely on the asymptotic behavior of $P^{(1)}_{l}(\cos\theta)/\sin\theta$ and $\partial P^{(1)}_{l}(\cos\theta)/\partial \theta$, which for fixed $\theta$ and $\ell\rightarrow\infty$ is given \cite{vandeHulst-book-1981,Turyshev-Toth:2019} as
{}
\begin{eqnarray}
\frac{P^{(1)}_\ell(\cos\theta)}{\sin\theta}
&=& \Big(\frac{2\ell}{\pi\sin^3\theta}\Big)^{\frac{1}{2}} \sin\Big((\ell+{\textstyle\frac{1}{2}})\theta-{\textstyle\frac{\pi}{4}}\Big)+{\cal O}(\ell^{-\textstyle\frac{3}{2}}),
\label{eq:pi-l*}\\
\frac{dP^{(1)}_\ell(\cos\theta)}{d\theta}
&=&  \Big(\frac{2\ell^3}{\pi\sin\theta}\Big)^{\frac{1}{2}} \cos\Big((\ell+{\textstyle\frac{1}{2}})\theta-{\textstyle\frac{\pi}{4}}\Big)+{\cal O}(\ell^{-\textstyle\frac{1}{2}}).
\label{eq:tau-l*}
\end{eqnarray}

With these approximations, in the region of geometric optics the function $\beta(r,\theta)$ from (\ref{eq:beta*1*}) takes the form:
{}
\begin{eqnarray}
\beta(r,\theta)&=&E_0\frac{ue^{ik(r+r_0+r_g\ln 4k^2rr_0)}}{ikr}
\sum_{\ell=kR_\odot^\star}^\infty \ell^{-1}
e^{i\big(2\sigma_\ell+\frac{\ell^2}{2k\tilde r}\big)}
\times\nonumber\\
&&\times\,
\Big\{
 \Big(\frac{2\ell^3}{\pi\sin\theta}\Big)^{\frac{1}{2}} \Big(1-\frac{\ell^2}{2k^2r^2}\Big)\cos\Big(\ell\theta-{\textstyle\frac{\pi}{4}}\Big)+\Big(\frac{2\ell}{\pi\sin^3\theta}\Big)^{\frac{1}{2}} \sin\Big(\ell\theta-{\textstyle\frac{\pi}{4}}\Big)
\Big\}.~~~~~~~
\label{eq:S1-v0s}
\end{eqnarray}

For large $\ell\gg1$, the first term in the curly brackets of (\ref{eq:S1-v0s}) dominates, so that this expression may be given as
{}
\begin{eqnarray}
\beta(r,\theta)&=&E_0\frac{ue^{ik(r+r_0+r_g\ln 4k^2rr_0)}}{ikr}
\sum_{\ell=kR_\odot^\star}^\infty
\Big(\frac{2\ell}{\pi\sin\theta}\Big)^{\frac{1}{2}} \Big(1-\frac{\ell^2}{2u^2k^2r^2}\Big)
e^{i\big(2\sigma_\ell+\frac{\ell^2}{2k\tilde r}\big)}
\cos\big(\ell\theta-{\textstyle\frac{\pi}{4}}\big).~~~~~~~
\label{eq:S1-v0s+}
\end{eqnarray}

To evaluate this expression, we again use the method of stationary phase. For this, representing (\ref{eq:S1-v0s+}) in the form of an integral over $\ell$, we have:
{}
\begin{eqnarray}
\beta(r,\theta)&=&-E_0\frac{ue^{ik(r+r_0+r_g\ln 4k^2rr_0)}}{kr}
\int_{\ell=kR_\odot^\star}^\infty \hskip-5pt   \frac{\sqrt{\ell}d\ell}{\sqrt{2\pi\sin\theta}} \Big(1-\frac{\ell^2}{2u^2k^2r^2}\Big)
e^{i\big(2\sigma_\ell+\frac{\ell^2}{2k\tilde r}\big)}
\Big(e^{i(\ell\theta+{\textstyle\frac{\pi}{4}})}-e^{-i(\ell\theta+{\textstyle\frac{\pi}{4}})}\Big).~~~~~~~
\label{eq:S1-v0s+int*}
\end{eqnarray}

Expression (\ref{eq:S1-v0s+int*}) shows that the $\ell$-dependent part of the phase has a structure identical to that of (\ref{eq:S-l}). Therefore, the same solutions for the points of stationary phase apply. As a result, using (\ref{eq:S-l-pri}) and (\ref{eq:sf*}), from (\ref{eq:S1-v0s+int*}) and for the first family of solutions (\ref{eq:S-l-pri}) or $\ell_{\tt in}=kb_{\tt in}$, to the order of ${\cal O}(\theta^4)$, we have the amplitude in the stationary phase solution
{}
\begin{eqnarray}
\frac{\sqrt{\ell_{\tt in}}}{\sqrt{2\pi\sin\theta}}\Big(1-\frac{\ell^2_{\tt in}}{2u^2k^2r^2}\Big)\sqrt{\frac{2\pi}{\varphi''(\ell_{\tt in})}}&=&
\sqrt{\mp1}k\tilde ru^{-1}\Big(\cos(\theta+\beta)-\frac{r_g}{r}\Big).~~~
\label{eq:S-l3p}
\end{eqnarray}

As a result, similarly to (\ref{eq:Pi_s_exp4+1*}), the expression for the $\beta_{\tt in}(r,\theta)$  takes the form (with $\varphi_{\tt in}(r,\theta)$ is given by (\ref{eq:phase_rr0}))
{}
\begin{eqnarray}
\beta_{\tt in}(r,\theta)&=&
E_0u^{-1}\frac{r_0}{r+r_0}\Big(\cos(\theta+\beta)-\frac{r_g}{r}\Big) e^{ik\varphi_{\tt in}(r,\theta)}
+{\cal O}(\theta^4, \frac{r_g}{r}\theta^2).~~~~~
\label{eq:Pi_s_exp4+1pp}
\end{eqnarray}

Now we turn our attention to the second family of solutions in (\ref{eq:S-l-pri}) or for $\ell_{\tt s}=kb_{\tt s}$. Similarly to (\ref{eq:S-l3p+*=}), we have
{}
\begin{eqnarray}
\frac{\sqrt{\ell_{\tt s}}}{\sqrt{2\pi\sin\theta}} \Big(1-\frac{\ell^2_{\tt s}}{2u^2k^2r^2}\Big)\sqrt{\frac{2\pi}{\varphi''(\ell_{\tt s})}}&=&
\sqrt{\pm1} \frac{2kr_g}{\theta^2}+{\cal O}(\theta^2,r^2_g),~~~~
\label{eq:S-l3p+*=2}
\end{eqnarray}
which yields the following result for $\beta_{\tt s}(r,\theta)$:
{}
\begin{eqnarray}
\beta_{\tt s}(r,\theta)&=&
E_0\frac{r_g}{2r\sin^2\textstyle{\frac{1}{2}}(\theta+\beta)}
e^{ik\big(r_0+r_g\ln 2kr_0+r+r_g\ln kr(1-\cos(\theta+\beta))\big)+2i\sigma_0}
+{\cal O}(\theta^2, \frac{r_g}{r}\theta^2).~~~~~
\label{eq:beta-2}
\end{eqnarray}

\subsection{Evaluating the  function $\gamma(r,\theta)$}
\label{sec:amp_func-der}

To determine the remaining components of the EM field (\ref{eq:DB-sol00p*}), we need to evaluate  the behavior of the function $\gamma(r,\theta)$ from  (\ref{eq:gamma*1*}). For that, we use the asymptotic behavior of $P^{(1)}_{l}(\cos\theta)/\sin\theta$ and $\partial P^{(1)}_{l}(\cos\theta)/\partial \theta$ from (\ref{eq:pi-l*}) and (\ref{eq:tau-l*}), and rely on the method of stationary phase.
Similarly to (\ref{eq:S1-v0s}), we drop the second term in the curly brackets in (\ref{eq:gamma*1*}). The resulting expression for $\gamma(r, \theta)$, for large partial momenta, $\ell\gg1$,  is now determined from the following integral:
{}
\begin{eqnarray}
\gamma(r, \theta)&=& E_0\frac{ue^{ik(r+r_0+r_g\ln 4k^2rr_0)}}{kr} \int_{\ell=kR_\odot^\star}^\infty \hskip -5pt
\frac{\sqrt{\ell}d\ell}{\sqrt{2\pi\sin\theta}}e^{i\big(2\sigma_\ell+\frac{\ell^2}{2k\tilde r}\big)}
\Big(e^{i(\ell\theta+{\textstyle\frac{\pi}{4}})}-e^{-i(\ell\theta+{\textstyle\frac{\pi}{4}})}\Big).
  \label{eq:gamma**1*}
\end{eqnarray}

Clearly, this expression yields the same points of stationary phase (\ref{eq:S-l}), thus, all the  results obtained in Sec.~\ref{sec:radial-comp} are also relevant here. Therefore,  the $\ell$-dependent amplitude of (\ref{eq:gamma**1*}), for the first family of solutions (\ref{eq:S-l-pri}), $\ell_{\tt in}=kb_{\tt in}$, is evaluated to be
{}
\begin{equation}
\frac{\sqrt{\ell_{\tt in}}}{\sqrt{2\pi\sin\theta}}\sqrt{\frac{2\pi}{\varphi''(\ell_{\tt in})}}=\pm\sqrt{\mp1}k\tilde r+{\cal O}(\theta^3, \frac{r_g}{r}\theta^2).
\label{eq:S-l3p+0*}
\end{equation}

With $\varphi_{\tt in}(r,\theta)$ from (\ref{eq:phase_rr0}),  the function $ \gamma_{\tt in}(r,\theta)$ is given as
{}
\begin{eqnarray}
 \gamma_{\tt in}(r,\theta)&=&
E_0u \frac{r_0}{r+r_0}e^{i\varphi_{\tt in}(r,\theta)}+{\cal O}(\theta^3,\frac{r_g}{r}\theta^2).~~~~~
\label{eq:Pi_s_exp4+1gg}
\end{eqnarray}

Finally, for the second family of solutions (\ref{eq:S-l-pri}), $\ell_{\tt s}=kb_{\tt s}$, the result for $\gamma_{\tt s}(r,\theta)$ is identical to that given by (\ref{eq:beta-2}).

\subsection{Solution for the EM field in the region of geometric optics}
\label{sec:EM-fieldsol}

To determine the components of the entire incident EM field in the region of geometric optics, we use the expressions that were obtained for the functions $\alpha_{\tt in}(r,\theta)$, $\beta_{\tt in}(r,\theta)$ and $\gamma_{\tt in}(r,\theta)$, which are given by (\ref{eq:Pi_s_exp4+1*}), (\ref{eq:Pi_s_exp4+1pp}) and (\ref{eq:Pi_s_exp4+1gg}), correspondingly, and substitute them in (\ref{eq:DB-sol00p*}).
Before we present the resulting solution for the EM field, we recognize that these functions $\alpha_{\tt in}(r,\theta)$, $\beta_{\tt in}(r,\theta)$ and $\gamma_{\tt in}(r,\theta)$ are given with respect to the optical axis (as evident from (\ref{eq:phase_rr0-vec})) that connects the point source with the center of the Sun. However, these functions are still constrained to the plane $\phi=\phi_0$. In order to regain the axial symmetry, we need to rotate the solution (\ref{eq:DB-sol00p*}) by the angle $\beta$ in the plane defined by $\phi_0$. As we mentioned earlier, because $\beta$ is very small, any contribution of such a rotation to the amplitude to the EM field is negligibly small. After performing the needed substitutions and implementing the rotation by $\beta$, we establish the solution for the incident wave produced by the Debye potential $\Pi_0$ from (\ref{eq:Pi-s_a+0}):
{}
\begin{eqnarray}
  \left( \begin{aligned}
{   D}_r& \\
{   B}_r& \\
  \end{aligned} \right)_{\tt \hskip -4pt in}&=& \frac{E^{\tt s}_0u^{-1}}{r+r_0}
\sin(\theta+\beta)\Big(1+\frac{r_g}{r(1-\cos(\theta+\beta))}\Big)
\left( \begin{aligned}
\cos\overline\phi\\
\sin\overline\phi\\
  \end{aligned} \right)
  e^{i \big(\varphi_{\tt in}(r,\theta)-\omega t\big)},
  \label{eq:DB-t-pl=p10} \\
    \left( \begin{aligned}
{   D}_\theta& \\
{   B}_\theta& \\
  \end{aligned} \right)_{\tt \hskip -4pt in} &=&
  \frac{E^{\tt s}_0u^{-1}}{r+r_0}
\Big(\cos(\theta+\beta)-\frac{r_g}{r}\Big)
\left( \begin{aligned}
\cos\overline\phi\\
\sin\overline\phi\\
  \end{aligned} \right)  e^{i \big(\varphi_{\tt in}(r,\theta)-\omega t\big)},
   \label{eq:DB-th=p10} \\
   \left( \begin{aligned}
{   D}_\phi& \\
{   B}_\phi& \\
  \end{aligned} \right)_{\tt \hskip -4pt in} &=&
  \frac{E^{\tt s}_0u}{r+r_0}
  \left( \begin{aligned}
-\sin\overline\phi\\
\cos\overline\phi\\
  \end{aligned} \right) \, e^{i \big(\varphi_{\tt in}(r,\theta)-\omega t\big)},
  \label{eq:DB-t-pl=p20}
\end{eqnarray}
where the phase $\varphi_{\tt in}(r,\theta)$ is given by (\ref{eq:phase_rr0}) or, equivalently, by (\ref{eq:phase_rr0-vec}).

This is the solution for the EM field in the geometric optics region formed by the solar gravitational monopole.

\subsection{Solution for the EM field in the weak interference region}
\label{sec:EM-field-sol-wi}

We recall (see \cite{Turyshev-Toth:2017} for details) that in the case of gravitational scattering, there are two waves that characterize the overall scattering process in the region of weak interference: the incident wave given by (\ref{eq:DB-t-pl=p10})--(\ref{eq:DB-t-pl=p20}) and the scattered wave with $\alpha_{\tt s}(r,\theta)$ and $\beta_{\tt s}(r,\theta)=\gamma_{\tt s}(r,\theta)$ are given by (\ref{eq:alpha-2}), (\ref{eq:beta-2}), correspondingly, leading to the following form of the scattered wave with $(D_r, B_r)_{\tt \hskip 0pt s}= {\cal O}(r^2_g)$ and the rest of the components given as
{}
\begin{eqnarray}
    \left( \begin{aligned}
{   D}_\theta& \\
{   B}_\theta& \\
  \end{aligned} \right)_{\tt \hskip -4pt s} =   \left( \begin{aligned}
{   B}_\phi& \\
-{   D}_\phi& \\
  \end{aligned} \right)_{\tt \hskip -4pt s}= \frac{E^{\tt s}_0}{r+r_0}
 \frac{r_g}{2r\sin^2\frac{1}{2}(\theta+\beta)}
 \left( \begin{aligned}
\cos\overline\phi\\
\sin\overline\phi\\
  \end{aligned} \right)
  e^{i\big(k\big(r_0+r_g\ln 2kr_0+r+r_g\ln kr(1-\cos(\theta+\beta))\big)+2\sigma_0-\omega t \big)}.
   \label{eq:scat-th-tot}
\end{eqnarray}

As a result, we established the EM field in the region of weak interference in the presence of the incident wave and the scattered wave, given by (\ref{eq:DB-t-pl=p10})--(\ref{eq:DB-t-pl=p20}) and  by (\ref{eq:scat-th-tot}), correspondingly.

Note that the way we handled the sums in (\ref{eq:alpha*1*})--(\ref{eq:gamma*1*})---replacing them throughout this section with integrals over $\ell$ and then evaluating the integrals via the method of a stationary phase---amounts to an integral in the lens plane typically encountered in the models for weak gravitational lensing. In fact, results (\ref{eq:DB-t-pl=p10})--(\ref{eq:DB-t-pl=p20}) and (\ref{eq:scat-th-tot}) may now be used to compute the energy flux at the image plane, similarly to that done in Sec.~II.F of \cite{Turyshev-Toth:2017}.  Although this is a rather simple step technically, we will not discuss it here as such a development is beyond the scope of the present paper.

This completes the derivation for the EM field in the geometric optics and  weak interference regions formed by the solar gravitational monopole. We now turn out attention to the strong interference region, which is the region of greatest importance for imaging with the SGL.

\section{EM field in the strong interference region}
\label{sec:IF-region}

We are interested in the area behind the Sun, reachable by light rays with impact parameters $b\geq R_\odot^\star$. The focal region of the SGL begins where $r\geq b^2/2r_g$ and  $0\leq \theta\simeq \sqrt{2r_g/r}$. The EM field in this region is derived from the Debye potential (\ref{eq:Pi-s_a+0}) and is given by the factors $\alpha(r,\theta)$, $\beta(r,\theta)$ and $\gamma(r,\theta)$ from (\ref{eq:alpha*1*})--(\ref{eq:gamma*1*}), which we now calculate.

\subsection{The function $\alpha(r,\theta)$ and the radial components of the EM field}
\label{sec:alpha-IF}

We again begin with the investigation of $\alpha(r,\theta)$ from  (\ref{eq:alpha*1*}).
To evaluate this expression in the interference region where $0\leq \theta \simeq \sqrt{2r_g/r}$, we use the asymptotic representation for $P^{(1)}_l(\cos\theta)$ from \cite{Bateman-Erdelyi:1953,Korn-Korn:1968,Kerker-book:1969}, valid when $\ell\to\infty$:
{}
\begin{eqnarray}
P^{(1)}_\ell(\cos\theta)&=& \frac{\ell+{\textstyle\frac{1}{2}}}{\cos{\textstyle\frac{1}{2}}\theta}J_1\big((\ell+{\textstyle\frac{1}{2}})2\sin{\textstyle\frac{1}{2}}\theta\big).
\label{eq:pi-l0}
\end{eqnarray}

We use the approximation above and replace the sum in the resulting expression
(\ref{eq:alpha*1*}) with an integral to be evaluated with the method of stationary phase:
 {}
\begin{eqnarray}
\alpha(r,\theta) &=& -E_0\frac{ue^{ik(r+r_0+r_g\ln 4k^2rr_0)}}{k^2r^2}\int_{\ell=kR^\star_\odot}^\infty  \ell^2 d\ell e^{i\big(2\sigma_\ell+\frac{\ell^2}{2k\tilde r}\big)}\Big(1-\frac{\ell^2}{4u^2k^2r^2} \Big)J_1\big(\ell \theta\big)
+{\cal O}\big(r^2_g,(kr)^{-3},\theta^2\big).
\label{eq:alpha*IF*int}
\end{eqnarray}

We see that the $\ell$-dependent phase in this expression is given as
{}
\begin{eqnarray}
\varphi(\ell)&=& 2\sigma_\ell+\frac{\ell^2}{2k\tilde r}+{\cal O}((kr)^{-3})=-2kr_g\ln \ell+\frac{\ell^2}{2k\tilde r}+{\cal O}((kr)^{-3}).
\label{eq:IF-phase}
\end{eqnarray}

The phase is stationary when $d\varphi(\ell)/d\ell=0$, resulting in
{}
\begin{eqnarray}
-\frac{2kr_g}{\ell}+\frac{\ell}{k\tilde r}={\cal O}((kr)^{-3})\qquad \Rightarrow\qquad
\ell^2 =2 k^2r_g\tilde r +{\cal O}((kr)^{-1}) \qquad {\rm or} \qquad \ell_0=k\sqrt{2r_g\tilde r}.
\label{eq:IF-phase3}
\end{eqnarray}
This solution, $\ell_0$, represents the smallest partial momenta for the light trajectories to reach a particular heliocentric distance, $r$, on the optical axis (Fig.~\ref{fig:geom-opt}). It is consistent with the solution to the equation for geodesics (see Appendix~B in \cite{Turyshev-Toth:2017}) which yields the solution for the impact parameter of $b=\sqrt{2r_g\tilde r}$. Note that we choose  $\ell$ to be positive.

We now return to  evaluating (\ref{eq:alpha*IF*int}). The solution given by (\ref{eq:IF-phase3}) allows us to compute the stationary phase (\ref{eq:IF-phase}):
{}
\begin{eqnarray}
\varphi(\ell_0)&=& -kr_g\ln 2k\tilde r+\sigma_0 +{\textstyle\frac{\pi}{2}},
\label{eq:IF-phase00}
\end{eqnarray}
where $\sigma_0$ is the constant given as $\sigma_0=\arg \Gamma(1-ikr_g)$ \cite{Morse-Feshbach:1953}. For large values of $kr_g\rightarrow\infty$ this constant is evaluated  as $\sigma_0=-kr_g\ln (kr_g/e)-\frac{\pi}{4}$ (see details in \cite{Turyshev-Toth:2017}).
Next, using (\ref{eq:IF-phase}), we compute the relevant
$\varphi''(\ell)={d^2\varphi_{}}/{d\ell^2} $ as
\begin{eqnarray}
\dfrac{d^2\varphi_{}}{d\ell^2} &=& \dfrac{1}{k\tilde r}+\frac{2kr_g}{\ell^2}
\qquad \Rightarrow \qquad
\sqrt{\frac{2\pi}{\varphi''(\ell_0)}}=\sqrt{\pi k\tilde r}.
\label{eq:IF-phase*6}
\end{eqnarray}

With (\ref{eq:IF-phase*6}), we now have the amplitude of the integrand in (\ref{eq:alpha*IF*int}), for $\ell$ from (\ref{eq:IF-phase3}), taking the form
{}
\begin{eqnarray}
\ell^2_0\Big(1-\frac{\ell^2_0}{4u^2k^2r^2}\Big)J_1\big(\ell \theta\big)\sqrt{\frac{2\pi}{\varphi''(\ell_0)}}&=&
k^22r_g\tilde r \sqrt{\pi k\tilde r}\Big(1-\frac{r_g\tilde r}{4r^2} +{\cal O}(r_g^2, (kr)^{-1})\Big)J_1\big(k\sqrt{2r_g\tilde r} \theta\big).
\label{eq:IF-phase*7}
\end{eqnarray}

As a result, the expression for $\alpha(r,\theta)$ from (\ref{eq:alpha*IF*int}) becomes
{}
\begin{eqnarray}
\alpha(r,\theta)&=&
-iE_0\sqrt{\frac{2r_g}{\tilde r} }\sqrt{2\pi kr_g}e^{i\sigma_0}\Big(\frac{r_0}{r+r_0}\big)^2J_1\big(k\sqrt{2r_g\tilde r} \theta\big)e^{ik\big(r+r_0+r_g\ln 2k(r+r_0)\big)}.
\label{eq:IF-phase*7*}
\end{eqnarray}

We can use the same approach to compute the remaining two factors $\beta(r,\theta)$ and $\gamma(r,\theta)$.

\subsection{The function $\beta(r,\theta)$ and the $\theta$-components of the EM field}
\label{sec:beta-IF}

Similarly to \cite{Turyshev-Toth:2017,Turyshev-Toth:2019}, to evaluate the magnitude of the function $\beta(r,\theta)$, we  need to establish the asymptotic behavior of the Legendre polynomials $P^{(1)}_{l}(\cos\theta)$ in the relevant regime. The asymptotic formulae for the Legendre polynomials  if $w=(\ell+{\textstyle\frac{1}{2}})\theta$ is fixed and $\ell\rightarrow \infty$ are \cite{vandeHulst-book-1981}:
{}
\begin{eqnarray}
\frac{P^{(1)}_\ell(\cos\theta)}{\sin\theta}&=& {\textstyle\frac{1}{2}}\ell(\ell+1)\Big(J_0(w)+J_2(w)\Big),
\label{eq:pi-l}
\qquad
\frac{dP^{(1)}_\ell(\cos\theta)}{d\theta}= {\textstyle\frac{1}{2}}\ell(\ell+1)\Big(J_0(w)-J_2(w)\Big).
\label{eq:tau-l}
\end{eqnarray}

Using (\ref{eq:tau-l}), we transform (\ref{eq:beta*1*}) by also replacing the sum in the resulting expression with an integral to derive $\beta$ as
 {}
\begin{eqnarray}
\beta(r,\theta) &=& E_0\frac{ue^{ik(r+r_0+r_g\ln 4k^2rr_0)}}{ikr}\int_{\ell=kR^\star_\odot}^\infty \ell d\ell e^{i\big(2\sigma_\ell+\frac{\ell^2}{2k\tilde r}\big)}
\Big\{J_0\big(\ell\theta\big)- {\textstyle\frac{1}{2}}\Big(J_0\big(\ell\theta\big)-J_2\big(\ell\theta\big)\Big)\frac{\ell^2}{2u^2k^2r^2} \Big\}.
  \label{eq:beta*IF*int}
\end{eqnarray}

As the $\ell$-dependent phase in (\ref{eq:beta*IF*int}) is the same as (\ref{eq:alpha*IF*int}), corresponding results from Sec.~\ref{sec:alpha-IF} are also applicable here. In fact, the same solutions for the points of stationary phase apply. As a result, using (\ref{eq:IF-phase3}), (\ref{eq:IF-phase*6}), from (\ref{eq:beta*IF*int}), we have
{}
\begin{eqnarray}
\ell_0\Big\{J_0\big(\ell_0\theta\big)- {\textstyle\frac{1}{2}}\Big(J_0\big(\ell_0\theta\big)-J_2\big(\ell_0\theta\big)\Big)
\frac{\ell^2_0}{2u^2k^2r^2} \Big\}\sqrt{\frac{2\pi}{\varphi''(\ell_0)}}=
k\tilde r\sqrt{2\pi kr_g}
\Big\{J_0\big(k\sqrt{2r_g\tilde r}\theta\big) +{\cal O}\big(\frac{r_g}{r}, r_g^2\big)\Big\}.~~~
\label{eq:beta*IF*=1}
\end{eqnarray}

As a result, the expression for $ \beta(r,\theta)$  in the interference region takes the form
{}
\begin{eqnarray}
 \beta(r,\theta)&=& E_0\sqrt{2\pi kr_g}e^{i\sigma_0}\frac{r_0}{r+r_0}
J_0\Big(k\sqrt{2r_g\tilde r}\theta\Big)e^{ik(r+r_0+r_g\ln 2k(r+r_0))}\Big(1+{\cal O}\big(\frac{r_g}{r}, r_g^2\big)\Big).
\label{eq:beta*IF*=2}
\end{eqnarray}

\subsection{The function $\gamma(r,\theta)$ and the $\phi$-components of the EM field}
\label{sec:gamma-IF}

The $\phi$-components of the EM field is govern by the factor  $\gamma(r,\theta)$ from (\ref{eq:gamma*1*}). Similarly to the discussion in Section \ref{sec:beta-IF}, we use (\ref{eq:pi-l*})--(\ref{eq:tau-l*}) to transform resulting expression for  $\gamma(r,\theta)$  to an integral, also taking $\ell\gg1$:
{}
\begin{eqnarray}
\gamma(r,\theta) &=& E_0\frac{ue^{ik(r+r_0+r_g\ln 4k^2rr_0)}}{ikr}\int_{\ell=kR^\star_\odot}^\infty\ell d\ell e^{i\big(2\sigma_\ell+\frac{\ell^2}{2k\tilde r}\big)}
\Big\{J_0(\ell\theta)-{\textstyle\frac{1}{2}}\Big(J_0(\ell\theta)+J_2(\ell\theta)\Big)\frac{\ell^2}{2u^2k^2r^2}\Big\}.
  \label{eq:gamma*IF*int}
\end{eqnarray}

As we noticed while deriving (\ref{eq:beta*IF*=1}), the term $\propto \ell^2/2u^2k^2r^2$ produces a contribution of ${\cal O}(r_g/r)$, which is negligible in the interference region. This term may also be neglected in the integrand of (\ref{eq:gamma*IF*int}). With this simplification, the resulting Eq.~(\ref{eq:gamma*IF*int}) is identical to that of (\ref{eq:beta*IF*int}). Therefore, we conclude that  in the interference region $\gamma(r,\theta)=\beta(r,\theta)+{\cal O}({r_g}/{r}, r_g^2)$.

\subsection{The EM field in the interference region}
\label{sec:amp_func-IF}

Now we are ready to present the components of the EM field in the interference region. We do that by using the expressions that we obtained for the functions $\alpha(r,\theta)$, $\beta(r,\theta)=\gamma(r,\theta)+{\cal O}(r_g/r)$, which are given by (\ref{eq:IF-phase*7*}), (\ref{eq:beta*IF*=2}), correspondingly, and substitute them in (\ref{eq:DB-sol00p*}). As a result, we establish the solution for the EM field produced by the Debye potential $\Pi$ given by (\ref{eq:Pi-s_a+0}) for the EM wave in the interference region.
After implementing, as before, the rotation by $\beta$ in the plane defined by $\phi_0$, we obtain the solution
 in the following form:
{}
\begin{eqnarray}
  \left( \begin{aligned}
{   D}_r& \\
{   B}_r& \\
  \end{aligned} \right) &=& -iE_0
\sqrt{\frac{2r_g}{\tilde r} }\sqrt{2\pi kr_g}e^{i\sigma_0}\Big(\frac{r_0}{r+r_0}\Big)^2J_1\Big(k\sqrt{2r_g\tilde r}
(\theta+\beta)
\Big)e^{i\big(k(r+r_0+r_g\ln2k(r+r_0))-\omega t\big)}\left( \begin{aligned}
\cos\overline\phi\\
\sin\overline\phi\\
  \end{aligned} \right),~~~~~
  \label{eq:IF-DBr0} \\
    \left( \begin{aligned}
{   D}_\theta& \\
{   B}_\theta& \\
  \end{aligned} \right) &=&   \left( \begin{aligned}
{   B}_\phi& \\
-{   D}_\phi& \\
  \end{aligned} \right)= E_0
\sqrt{2\pi kr_g}e^{i\sigma_0}\frac{r_0}{r+r_0}J_0\Big(k\sqrt{2r_g\tilde r}
(\theta+\beta)
\Big)e^{i\big(k(r+r_0+r_g\ln 2k(r+r_0))-\omega t\big)}\left( \begin{aligned}
\cos\overline\phi\\
\sin\overline\phi\\
  \end{aligned} \right).
   \label{eq:IF-DBth0}
\end{eqnarray}

The radial component of the EM field (\ref{eq:IF-DBr0}) is negligibly small compared to the other two components, which is consistent with the fact that while passing through solar gravity the EM wave preserves its transverse structure.

Eqs.~(\ref{eq:IF-DBr0})--(\ref{eq:IF-DBth0}) describe the EM field in the interference region of the SGL  in the spherical coordinate system.  To study this field on the image plane, we follow the approach demonstrated in  \cite{Turyshev-Toth:2017}, where instead of spherical coordinates $(r,\theta,\phi)$, we introduced a cylindrical coordinate system $(\rho,\phi,z)$, more convenient for these purposes.  In the region $r \gg r_g$, this can be done by defining $R=ur = r+r_g/2+{\cal O}(r_g^2)$ and introducing the coordinate transformations $ \rho=R\sin(\theta+\beta),$ $ z=R\cos(\theta+\beta)$, which, from (\ref{eq:metric-gen}), result in the following line element:
{}
\begin{eqnarray}
ds^2&=&u^{-2}c^2dt^2-u^2\big(dr^2+r^2(d\theta^2+\sin^2\theta d\phi^2)\big)=u^{-2}c^2dt^2-\big(d\rho^2+(\rho-\beta z)^2d\phi^2+u^2dz^2\big)+{\cal O}(r_g^2, \beta^2).
\label{eq:cyl_coord}
\end{eqnarray}

\begin{figure}
\includegraphics[scale=0.35]{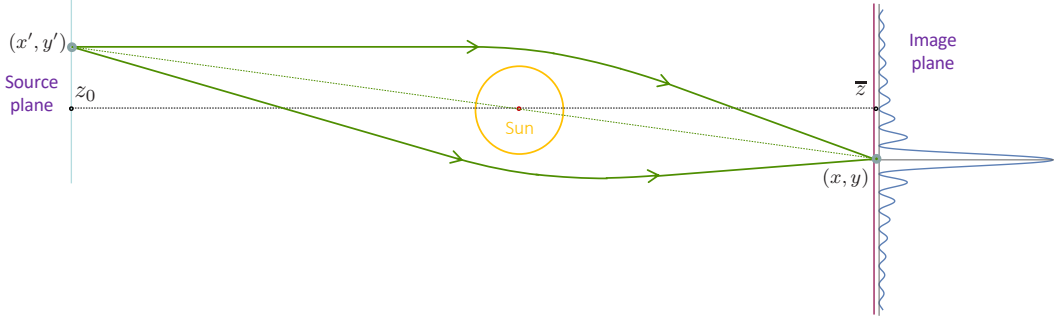}
\caption{\label{fig:single}The SGL maps a point source with coordinates $(x',y')$ in the source plane to a point with coordinates $(x,y)=-(z/z_0)(x',y')$ in the image plane. The rotation of the PSF pattern, evident in Fig.~\ref{fig:geom-opt}, is not emphasized here.
}
\end{figure}

As a result, using (\ref{eq:IF-DBr0})--(\ref{eq:IF-DBth0}), for a high-frequency EM wave (i.e., neglecting terms $\propto(kr)^{-1}$) and for $r\gg r_g$, we derive the  components of the EM field near the optical axis, which, together with (\ref{eq:tildebeta}) and up to terms of ${\cal O}(\rho^2/z^2, \beta)$, in the paraxial approximation (see discussion after (\ref{eq:DB-t-pl=p20})), take the form
{}
\begin{eqnarray}
    \left( \begin{aligned}
{E}_\rho& \\
{H}_\rho& \\
  \end{aligned} \right) =    \left( \begin{aligned}
{H}_\phi& \\
-{E}_\phi& \\
  \end{aligned} \right)&=&
    \frac{ {E}^{\tt s}_0}{r+r_0}
  \sqrt{2\pi kr_g}e^{i\sigma_0}
  J_0\Big(k\sqrt{2r_g\overline r}
   (\theta+\beta) \Big)
    e^{i\big(k(r+r_0+r_g\ln 2k(r+r_0))-\omega t\big)}
 \left( \begin{aligned}
\cos\overline\phi\\
\sin\overline\phi\\
  \end{aligned} \right),
  \label{eq:DB-sol-rho}
\end{eqnarray}
with the $z$-components of the EM wave behave as $({E}_z, {H}_z)\sim {\cal O}({\rho}/{z},\beta)$. The quantity $\overline r=r(1+r/r_0+{\cal O}(r^2/r_0^2))$ from (\ref{eq:r-bar}) denotes heliocentric distances along the line connecting the point source and the center of the Sun. Also, $r=\sqrt{z^2+\rho^2}=z+{\cal O}(\rho^2/z)$), $\theta=\rho/z+{\cal O}(\rho^2/z^2)$ and $\beta=b/z_0+{\cal O}(b^2/z_0^2)$. Note that these expressions were obtained  using the approximations (\ref{eq:pi-l}) and are valid for forward scattering when $\theta+\beta\approx 0$, or when $0\leq \rho\leq r_g$.

Using (\ref{eq:DB-sol-rho}), we now compute the energy flux on the image plane in the interference region of the SGL (see Fig.~\ref{fig:single}). The relevant components of the time-averaged Poynting vector for the EM field in the image volume, as a result, may be given as \cite{Turyshev-Toth:2017}:
{}
\begin{eqnarray}
{\bar S}_z&=&\frac{c}{8\pi}
\Big(\frac{E^{\tt s}_0}{r+r_0}\Big)^2
\frac{2\pi kr_g}{1-e^{-2\pi k r_g}}
J^2_0\Big(k\sqrt{2r_g \overline r}
   (\theta+\beta)\Big),
\label{eq:S_z*6z}
\end{eqnarray}
with ${\bar S}_\rho= {\bar S}_\phi=0$ for any practical purposes. Therefore, the non-vanishing component of the light amplification vector $ {\vec \mu}$, defined as ${\vec \mu}={\vec {\bar S}}/|{\vec{\bar S}}_0|$ (where $|{\bar {\vec S}}_0|=(c/8\pi)(E_0^{\tt s}/(r+r_0))^2$ is the time-averaged Poynting vector of a plane wave propagating in empty spacetime) takes the form:
{}
\begin{eqnarray}
{\bar \mu}_z&=&
\mu_0 J^2_0\Big(\frac{2\pi}{\lambda}\sqrt{2r_g \overline r}
   (\theta+\beta)
\Big),\qquad {\rm with}
\qquad
\mu_0=\frac{4\pi^2}{1-e^{-4\pi^2 r_g/\lambda}}
\frac{r_g}{\lambda},
\label{eq:S_z*6z-mu}
\end{eqnarray}
where $\mu_0$ is the light amplification factor for imaging a point source with observer positioned on  the optical axis that is set by position of that source and the origin of the heliocentric coordinate system. The amplification reaches its maximum value when the argument of the Bessel function vanishes which is happening when, for a given value of $\beta$ (that is set by the position of the point source), the angle $\theta$ takes opposite value, namely $\theta=-\beta$. Clearly, one recovers (\ref{eq:mu-point}) by taking in (\ref{eq:S_z*6z-mu})  the limit of $z_0\rightarrow \infty$ or, equivalently, by setting $\beta=0$.

The result given by Eq.~(\ref{eq:S_z*6z-mu}) can be used to study the image formation process in the case of an extended source.

\section{Image formation for an extended source with the SGL}
\label{sec:image_form}

\begin{figure}
\includegraphics[scale=0.38]{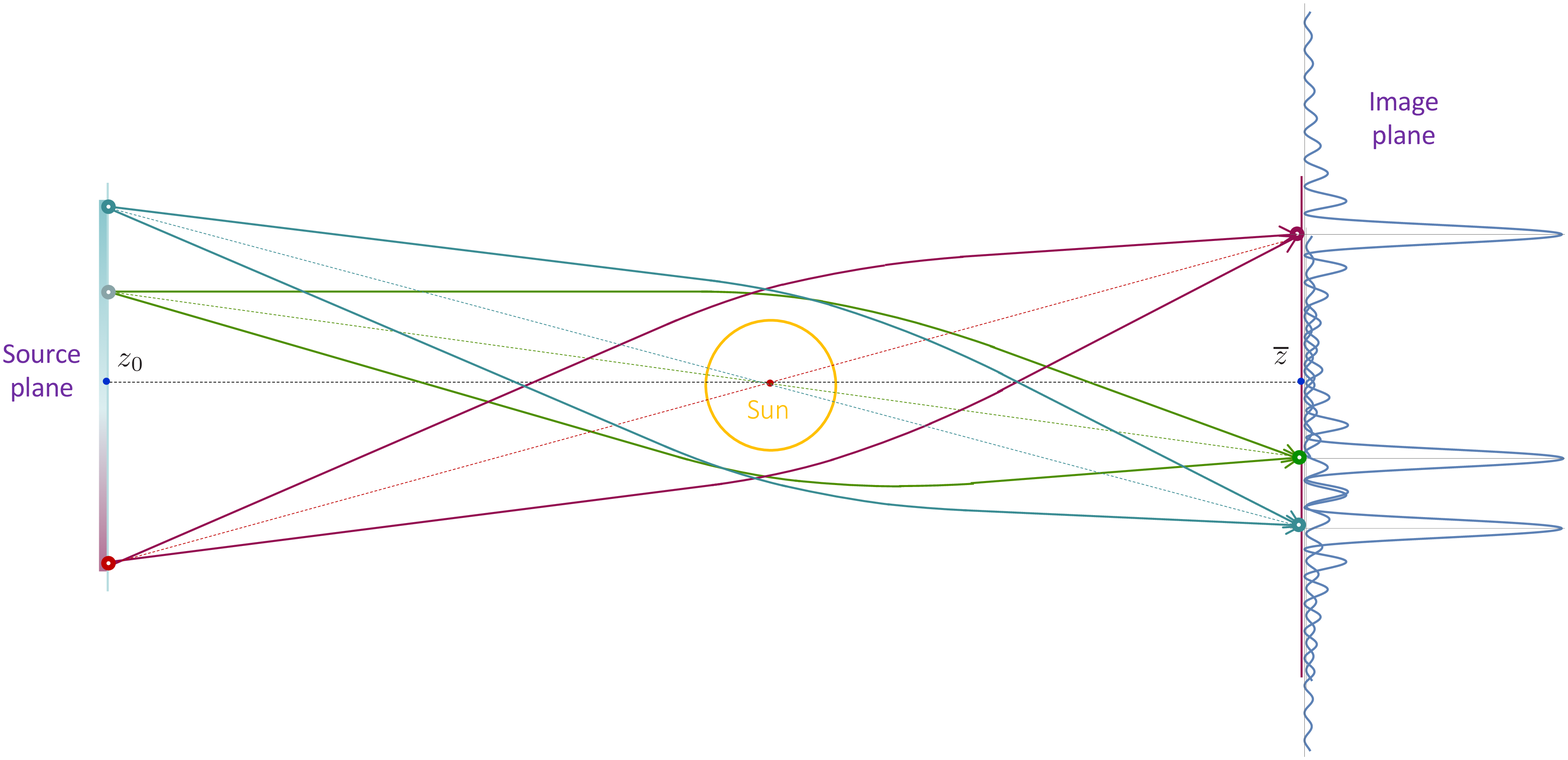}
\caption{\label{fig:multi}When the SGL maps an extended source (represented as multiple point sources) from the source to the image plane, the result is a set of overlapping PSF patterns. The center of each PSF represents the location in the image plane where an observer, looking back at the Sun, would see a complete Einstein ring attributed to that specific point source. Other point sources contribute light in the form of partial Einstein rings. $z_0$ and $\overline z$ denote heliocentric distances to the source plane and image plane, correspondingly. As in Fig.~\ref{fig:single}, rotation of the individual PSF patterns, evident in Fig.~\ref{fig:geom-opt}, is not emphasized here. }
\end{figure}

To study how the SGL forms an image of an extended source, we model that extended source as a collection of point sources in the source plane. We  integrate the point-spread function (PSF) for a single point source (\ref{eq:S_z*6z-mu}) over the extended source (Fig.~\ref{fig:multi}).

\subsection{Generalization to the case of an extended source}
\label{sec:SGL-gen-ext}

To discuss image formation in the case of an extended source, we introduce  the vector of position of an imaging telescope in the coordinate system corresponding to the optical axis, which can be done as $\overline{\vec r}=\overline r( {\vec n}+{\vec k}_\beta)+{\cal O}(\beta^2)$. Following \cite{Turyshev-Toth:2017}, we recognize that for small angles $\theta$ and $\beta$ the argument of the Bessel function in (\ref{eq:S_z*6z}) to the order of ${\cal O}(\theta^2,\beta^2)$ has the form
{}
\begin{eqnarray}
k\sqrt{2r_g \overline r} \big(\theta+\beta\big)&=&
2\sqrt{kr_g(k\overline r-({\vec k}\cdot{\overline {\vec  r}}))}= k\sqrt{2r_g\overline r}|{\vec k} \times({\vec k}\times({\vec n}+{\vec k}_\beta))|
=k\sqrt{2r_g\overline r}|{\vec n}_\perp+{\vec k}_{\beta\perp}|
=\nonumber\\
&=&k\sqrt{2r_g\overline r}
\sqrt{\Big(\frac{x}{\overline r}+\frac{b_x}{{ r}_0}\Big)^2+\Big(\frac{y}{{\overline r}}+\frac{b_y}{{ r}_0}\Big)^2}=k\sqrt{2r_g\overline r}\,\Big|\frac{\vec x}{\overline r}+\frac{\vec x'}{{ r}_0}\Big|,
\label{eq:arg}
\end{eqnarray}
where ${\vec n}_\perp=({x/{\overline r},y/{\overline r},0)}$, ${\vec k}_{\beta\perp}=(b_x/{ r_0},b_y/{ r_0},0)$ are the components of these two vectors perpendicular to $\vec k$.

We introduce the coordinate system where the $z$-axis lies on the principal optical axis, which is the line that connects the center of the exoplanet and the center of the Sun, before intersecting the image plane (see Fig.~\ref{fig:single}). It is convenient to express the results in terms of the distances along the principal optical axis, $z$ and $z_0$, rather than the distances along a particular optical axis $r$ and $r_0$.  In addition, in this coordinate system, any point in the source and/or image planes is described by a deviation from that principal optical axis and corresponding angles, namely $(x',y',z_0) \leftrightarrow (\rho',\phi',z_0)$ for the source and $(x,y,z) \leftrightarrow (\rho,\phi,z)$ for the image, correspondingly.

Using the new coordinate system, the distances in (\ref{eq:arg}) are related to each other as  $r=\sqrt{z^2+\rho^2}= z+{\cal O}(\rho^2/z)$, and $r_0=\sqrt{z_0^2+\rho'^2}= z_0+{\cal O}({\rho'^2}/{z_0})$, where we  neglected terms that are second order in $\rho$ and $\rho'$, as ${\rho}/{z}\ll1$ and ${\rho'}/{z_0}\ll 1$. As a result, the distance $\overline r$ is expressed as  $\overline r\rightarrow \overline z=z(1+z/z_0)$.

Now we can describe the imaging of an extended source. For that we can generalize (\ref{eq:S_z*6z-mu}) by using (\ref{eq:arg}), where, relying on the axial symmetry of the problem,  we replace one point on the surface of the source $(b_y,b_y,r_0)$ with a generic point with coordinates $(x',y',z_0)$. This yields the PSF of the SGL, expressed as a function of the location $\vec{x}$ of a point source at a finite distance from the Sun, and a location $\vec{x'}$ in the image plane:
{}
\begin{eqnarray}
{\bar \mu}_z({\vec x},{\vec x}')&=&
\mu_0J^2_0\Big(\frac{2\pi}{\lambda}
\sqrt{\frac{2r_g}{\overline z}}
|{\vec x}+\frac{\overline z}{{ z}_0}{\vec x'}|\Big).
\label{eq:S_z*6z-mu2}
\end{eqnarray}

This is our main result. It allows one to study the image formation process, develop realistic imaging scenarios, and perform relevant simulations.
The result (\ref{eq:S_z*6z-mu2}) is different form the one that was obtained earlier \cite{Turyshev-Toth:2017} and shown by (\ref{eq:mu-point}). The new result  explicitly accounts for the fact that the distance to the source is finite.  In addition, this expression also explicitly depends on the coordinates ${\vec x}$ and ${\vec x'}$, not restricting them to a plane as was done in (\ref{eq:mu-point}) and (\ref{eq:S_z*6z-mu}).

\subsection{Photometric imaging of an extended  source at a finite distance}
\label{sec:SGL-imaging-ext}

To produce an image of an astronomical source, the sources are assumed to be  non-coherent. This allows us to use photometric imaging techniques. In this case, a telescope is used as a ``light bucket'', measuring the total brightness of the Einstein ring at various locations in the image plane, corresponding to different parts of the source (see Fig.~\ref{fig:single}).

For an extended luminous source with the surface brightness of $B(x',y')$, the power density, $I_0(x,y)$, received on the image plane at a distance of $\overline z+z_0$ from the source  (see Fig.~\ref{fig:multi}) is computed by integrating the PSF (\ref{eq:S_z*6z-mu2}) over the surface of the extended source, which may be expressed as
\begin{eqnarray}
I_0(x,y)&=&\frac{\mu_0}{4\pi (\overline z+z_0)^2}\iint\displaylimits_{-\infty}^{+\infty}dx'dy'\, B(x',y')\,
J^2_0\Big(\frac{2\pi}{\lambda}\sqrt{\frac{2r_g}{\overline z}}
|{\vec x}+\frac{\overline z}{z_0}{\vec x}'|\Big),
\label{eq:power_dens}
\end{eqnarray}
where $B(x',y')$ is a function with compact support having non-zero values only within the source's dimensions.

Examining (\ref{eq:power_dens}), we see that monopole gravitational lens acts as a convex lens by focusing light, according to
{}
\begin{equation}
x=-\frac{\overline z}{z_0}x', \qquad y=-\frac{\overline z}{z_0}y'.
\end{equation}
This expression implies that the lens focuses light in the opposite quadrant in the image plane by also compressing the projected size of the source by a factor of ${z}/{z_0}\sim1.0\times 10^{-4}\,(z/650 ~{\rm AU}) (30~{\rm pc}/z_0)$. Thus, the diameter of the projection of an Earth-like exoplanet at those distances is reduced to $r_\oplus=R_\oplus (z/z_0)=1.34 (z/650 ~{\rm AU}) (30~{\rm pc}/z_0)$~km.

Given the image radius of $r_\oplus$, a telescope with aperture $d<2r_\oplus$  centered at a particular point $(x_0,y_0)$ on the image plane will receive the signal $P_d(x_0,y_0)=\iint dxdy \,I_0\big(x_0+x,y_0+y\big),$ where the integration is done within the telescope's aperture $|\vec x|\leq d/2$, and with $|{\vec x}_0+{\vec x}|\leq r_\oplus$, yielding the following result:
{}
\begin{eqnarray}
P_d(x_0,y_0)&=&\frac{\mu_0}{4\pi (\overline z+z_0)^2}\hskip -5pt\iint\displaylimits_{|{\vec x}|^2\leq (\frac{1}{2}d)^2}\hskip -5pt dxdy \iint\displaylimits_{-\infty}^{+\infty}dx'dy'\, B(x',y')\,
J^2_0\Big(\frac{2\pi}{\lambda}\sqrt{\frac{2r_g}{\overline z}}
|{\vec x}_0+{\vec x}+\frac{\overline z}{z_0}{\vec x}'|\Big).
\label{eq:power_rec2}
\end{eqnarray}

We note that although this result does not include the contribution from plasma in the solar corona \cite{Turyshev-Toth:2018-plasma,Turyshev-Toth:2019}, such a contribution may be easily incorporated if needed.  This may be the case if we were to use the SGL for an application at microwave frequencies where the effect of solar plasma is significant. For optical and IR wavelengths the effect of the solar plasma on the optical properties of the SGL is negligible.

This is our main result that may be used to study the image formation for an extended source. This result opens the way for using the SGL for imaging of faint targets positioned at a large but finite distance from the Sun.

Expression (\ref{eq:power_rec2}) is rather complex and must be evaluated numerically. However, an interesting limiting case exists that may still be treated analytically: that of a point source at a finite distance. For an incoherent source, the power received by the telescope is given by (\ref{eq:power_rec2}). For a point source positioned at the optical axis,  ${\vec x}'=0$ and the telescope at the center of the image,  ${\vec x}_0=0$, integral (\ref{eq:power_rec2}) is easy to evaluate analytically. Assuming, a uniform source brightness, $B(x',y')=B_{\tt s}$, we integrate (\ref{eq:power_rec2}):
{}
\begin{eqnarray}
P^0_d(0,0)&=&
\frac{\pi R_\oplus^2d^2B_{\tt s}}{16(\overline z+z_0)^2}\mu_0\Big\{ J^2_0\Big(\pi\frac{d}{\lambda}\sqrt{\frac{2r_g}{\overline z}}\Big)+J^2_1\Big(\pi\frac{d}{\lambda}\sqrt{\frac{2r_g}{\overline z}}\Big)\Big\}\approx
\frac{\pi R_\oplus^2dB_{\tt s}}{4(\overline z+z_0)^2}\sqrt{2r_g \overline z},
\label{eq:mu_av}
\end{eqnarray}
where we used the approximations for the Bessel functions for large arguments \cite{Abramovitz-Stegun:1965}:
\begin{eqnarray}
J_0(x)\simeq \sqrt{\frac{2}{\pi x}}\cos(x-{\textstyle\frac{\pi}{4}})+{\cal O}\big(x^{-1}\big)
\qquad {\rm and} \qquad
J_1(x)\simeq \sqrt{\frac{2}{\pi x}}\sin(x-{\textstyle\frac{\pi}{4}})+{\cal O}\big(x^{-1}\big),
\label{eq:BF}
\end{eqnarray}
which is appropriate for typical parameters relevant to imaging with the SGL.

Expression (\ref{eq:mu_av}) may be used to estimate the signal received from a distant unresolved source. Some approximations exist to allow for a semi-analytical treatment of extended sources; however, their description is out of the scope of this paper and will be given elsewhere.

\subsection{Image formation by an optical telescope at the image plane}
\label{sec:image-form-Fourier}

A classical imaging technique relies on an imaging telescope with a large focal plane sensor array capable of capturing high-resolution images. This technique is well developed for coherent sources; however, astronomical sources may not be treated as such.  Nevertheless, it is interesting to consider the image formation process with the SGL in the case when the extended source may be treated as coherent.

To produce images with the SGL, we represent an imaging telescope by a convex lens with focal distance $f$, and position the telescope at the image plane in the interference region \cite{Nambu:2013,Kanai-Nambu:2013,Nambu:2013b,Born-Wolf:1999}.  The amplitude of the wave just in front of the lens from (\ref{eq:DB-sol-rho}) and (\ref{eq:S_z*6z-mu2}) is given as
{}
\begin{eqnarray}
{\cal A}({\vec x},{\vec x}')&=&
  {E}_0  \sqrt{\mu_0}
  J_0\Big(k
\sqrt{\frac{2r_g}{\overline z}} |{\vec x}+\frac{\overline z}{{ z}_0}{\vec x}'|\Big).
  \label{eq:amp-w}
\end{eqnarray}

The presence of a convex lens with focal distance $f$ is equivalent to a Fourier transform of the wave (\ref{eq:amp-w}).  Consider ${\cal A}({\vec x},{\vec x}')$ to be the EM field just in front of the convex lens with focal distance of $\overline z=f$. We position the detector at the focal distance of the convex lens. Using the Fresnel-Kirchhoff diffraction formula, the wave's amplitude on the detector's focal plane at a pixel location of ${\vec p}=(x_i,y_i)$  is given by
{}
\begin{eqnarray}
{\cal A}({\vec p},{\vec x}')=\frac{i}{\lambda}\iint \displaylimits_{|{\vec x}|^2\leq (d/2)^2} \hskip -7pt  {\cal A}({\vec x},{\vec x}')e^{-i\frac{k}{2f}|{\vec x}|^2}\frac{e^{iks}}{s}d^2{\vec x},
  \label{eq:amp-w-f0}
\end{eqnarray}
where $s$ is the optical path.
The function $e^{-i\frac{k}{2f}|{\vec x}|^2}=e^{-i\frac{k}{2f}(x^2+y^2)}$ represents the action of the convex lens which transforms incident plane waves to spherical waves focusing at the focal point. Assuming that the focal length is sufficiently larger than the radius of the lens, we may approximate the optical path as $s=\sqrt{(x-x_i)^2+(y-y_i)^2+f^2}\sim f+\big((x-x_i)^2+(y-y_i)^2\big)/2f$. This allows us to present (\ref{eq:amp-w-f0}) as
{}
\begin{eqnarray}
{\cal A}({\vec p},{\vec x}')&=&
-  {E}_0  \sqrt{\mu_0}\frac{e^{ikf(1+{{\vec p}^2}/{2f^2})}}{i\lambda f}\iint\displaylimits_{|{\vec x}|^2\leq (\frac{1}{2}d)^2} d^2{\vec x}
  J_0\Big(k
\sqrt{\frac{2r_g}{\overline z}} |{\vec x}+\frac{\overline z}{{ z}_0}{\vec x}'|\Big) e^{-i\frac{k}{f}({\vec x}\cdot{\vec p})}.
  \label{eq:amp-w-f}
\end{eqnarray}

Integrating over the surface of the source, we have the amplitude of the  total EM field on the imaging detector:
{}
\begin{eqnarray}
{\cal A}({\vec p})&=&
\iint \displaylimits_{|{\vec x}'|^2\leq R_\oplus^2}\hskip -5pt {\cal A}({\vec p},{\vec x}')d^2{\vec x}'.
  \label{eq:amp-w-f3}
\end{eqnarray}

With the amplitude ${\cal A}({\vec p})$ given by (\ref{eq:amp-w-f3}), the EM field on the detector (denoted by subscript $\vec{p}$) is given as
{}
\begin{eqnarray}
    \left( \begin{aligned}
{E}_\rho& \\
{H}_\rho& \\
  \end{aligned} \right)_{\tt \hskip -3pt \vec p} =    \left( \begin{aligned}
{H}_\phi& \\
-{E}_\phi& \\
  \end{aligned} \right)_{\tt \hskip -3pt \vec p} &=&
  \frac{{\cal A}({\vec p})}{\overline z +z_0}
    e^{i\big(k(r+r_0+r_g\ln 2k(r+r_0))-\omega t\big)}
 \left( \begin{aligned}
 \cos\big(\phi +\phi_0\big)& \\
 \sin\big(\phi +\phi_0\big)& \\
  \end{aligned} \right).
  \label{eq:DB-sol-rho2}
\end{eqnarray}

Using these results, we compute the Poynting vector for the EM field emitted  by an extended source and received at a particular pixel ${\vec p}$ on the imaging detector. Given the form of EM field (\ref{eq:DB-sol-rho2}) and (\ref{eq:amp-w-f3}), the Poynting vector will have only one non-zero component, $S_z$. With overline and brackets denoting time averaging and ensemble averaging (over the oscillators on the source's surface), correspondingly, we compute $S_z$ as
 {}
\begin{eqnarray}
S_z({\vec p})=\frac{c}{4\pi}\big<\overline{[{\rm Re}{\vec E}\times {\rm Re}{\vec H}]}_z\big>=\frac{c}{8\pi}\frac{1}{(\overline z+z_0)^2}
\big<\big({\rm Re}{\cal A}({\vec p})e^{i\Omega(t)}\big)^2\big>,
  \label{eq:Pv}
\end{eqnarray}
where $\Omega(t)$ is the entire time-dependent phase given as $\Omega(t)=kf(1+{{\vec p}^2}/{2f^2})+k(r+r_0+r_g\ln 2k(r+r_0))-\omega t$.
The expectation value for $\big<\big({\rm Re}{\cal A}({\vec p})e^{i\Omega(t)}\big)^2\big>$ is given by the following expression:
 {}
\begin{eqnarray}
\big<\big({\rm Re}{\cal A}({\vec p})e^{i\Omega(t)}\big)^2\big>&=&\frac{{\mu_0}}{(\lambda f)^2}\bigg<{\rm Re}\, \Big(e^{i\Omega(t)}
\hskip -5pt
\iint \displaylimits_{|{\vec x}'_a|^2\leq R_\oplus^2} \hskip -5pt
d^2{\vec x}'_a
{E}_0({\vec x}'_a)   \hskip -5pt
 \iint\displaylimits_{|{\vec x}|^2\leq (d/2)^2}
  \hskip -5pt
  d^2{\vec x}\,
  J_0\Big(k
\sqrt{\frac{2r_g}{\overline r}} |{\vec x}+\frac{\overline r}{{\overline r}_0}{\vec x}_a'|\Big) e^{-i\frac{k}{f}({\vec x}\cdot{\vec p})}\Big) \times\nonumber\\
&&\hskip 25pt \times \, {\rm Re}\, \Big(e^{i\Omega(t)}\hskip -5pt
\iint \displaylimits_{|{\vec x}_b'|^2\leq R_\oplus^2} \hskip -5pt
d^2{\vec x}_b'
{E}_0({\vec x}_b')   \hskip -5pt
 \iint\displaylimits_{|{\vec x}|^2\leq (d/2)^2}
 \hskip -5pt d^2{\vec x}\,
  J_0\Big(k
\sqrt{\frac{2r_g}{\overline r}} |{\vec x}+\frac{\overline r}{{\overline r}_0}{\vec x}_b'|\Big) e^{-i\frac{k}{f}({\vec x}\cdot{\vec p})}\Big)\bigg>,
  \label{eq:Pv2}
\end{eqnarray}
where we introduced ${E}_0({\vec x}_a')$ and ${E}_0({\vec x}_b')$ to account for the fact that the extended source may be incoherent.

Expression (\ref{eq:Pv2}) is rather complex to be evaluated analytically in the general case. However, it can be evaluated for a point source on the optical axis, ${\vec x}'=0$. In this case, the integral in (\ref{eq:Pv2}) is easy to compute analytically. As we deal with the point source, we may model ${E}_0({\vec x}_a')={E}_0({\vec x}_b')=E_0\delta ({\vec x}')$, which takes care of the outer double integrals over $d^2 {\vec x}'$ in (\ref{eq:Pv2}). With this, we only need to integrate one inner double integral over $d^2 {\vec x}$. To do this, we introduce the relevant lens and detector coordinates as  $(x,y)=\rho(\cos\phi,\sin\phi)$, $(p_x,p_y)=\rho_i(\cos\phi_i,\sin\phi_i)$, correspondingly, and, using these new variables, we compute the integral $d^2 {\vec x}$ in (\ref{eq:Pv2}):
{}
\begin{eqnarray}
\int_0^{2\pi}\hskip-7pt d\phi \int_0^{d/2}\hskip-7pt\rho d\rho
  J_0\Big(k
\sqrt{\frac{2r_g}{\overline z}} \rho\Big) e^{-i\frac{k}{f}\rho \rho_i \cos(\phi-\phi_i)}&=&2\pi\int_0^{d/2}\hskip-7pt\rho d\rho
  J_0\Big(k
\sqrt{\frac{2r_g}{\overline z}} \rho\Big) J_0\Big(\frac{k \rho_i }{f}\rho\Big)=\nonumber\\
&&\hskip-150pt =\,
 \frac{\pi kd}{k^2(\frac{2r_g}{\overline z}-\frac{\rho_i^2}{f^2})} \Big\{
 \sqrt{\frac{2r_g}{\overline z}}J_0\Big({\textstyle\frac{1}{2}}kd \frac{\rho_i}{f}\Big)J_1\Big({\textstyle\frac{1}{2}}kd \sqrt{\frac{2r_g}{\overline z}}\Big)-  \frac{\rho_i}{f}J_0\Big({\textstyle\frac{1}{2}}kd\sqrt{\frac{2r_g}{\overline z}}\Big)J_1\Big({\textstyle\frac{1}{2}}kd \frac{\rho_i}{f}\Big)\Big\}.
  \label{eq:amp-w-f2d}
\end{eqnarray}

This result allows us to express the intensity of the EM signal at the detector.
After averaging over time, and using (\ref{eq:amp-w-f2d}) from (\ref{eq:Pv2}), we get the following components of the Poynting vector (NB: we arrange factors, e.g., of $d^2$ in order to present the Poynting vector in the form usually found in the literature \cite{Born-Wolf:1999}), to ${\cal O}({\overline z}/z_0)$:
{}
\begin{eqnarray}
S_z(\rho_i)=\frac{c}{8\pi}\frac{\mu_0E^{\tt s2}_0}{z_0^2}\Big(\frac{kd^2}{8  f}\Big)^2\bigg(
 \frac{2}{{\textstyle\frac{1}{2}}kd(\frac{2r_g}{\overline z}-\frac{\rho_i^2}{f^2})} \Big\{
 \sqrt{\frac{2r_g}{\overline z}}J_0\Big({\textstyle\frac{1}{2}}kd \frac{\rho_i}{f}\Big)J_1\Big({\textstyle\frac{1}{2}}kd \sqrt{\frac{2r_g}{\overline z}}\Big)-  \frac{\rho_i}{f}J_0\Big({\textstyle\frac{1}{2}}kd\sqrt{\frac{2r_g}{\overline z}}\Big)J_1\Big({\textstyle\frac{1}{2}}kd \frac{\rho_i}{f}\Big)\Big\}\bigg)^2.~
  \label{eq:amp-w-f2dp}
\end{eqnarray}

Taking a limit of $r_g \rightarrow 0$ in (\ref{eq:amp-w-f2dp}), we obtain
 the Poynting vector showing the classic Airy pattern:
{}
\begin{eqnarray}
S^0_z(\rho_i)
\big|_{r_g\rightarrow 0}=\frac{c}{8\pi}\frac{E^{\tt s2}_0}{(\overline z+z_0)^2}\Big(\frac{kd^2}{8f}\Big)^2\bigg(\frac{2J_1\big( {\textstyle\frac{1}{2}}kd \frac{\rho_i}{f}\big)}{{\textstyle\frac{1}{2}}kd\frac{\rho_i}{f}}\bigg)^2.
  \label{eq:amp-w-f2d2p}
\end{eqnarray}

\begin{figure}
\begin{center}
\includegraphics[width=0.25\linewidth]{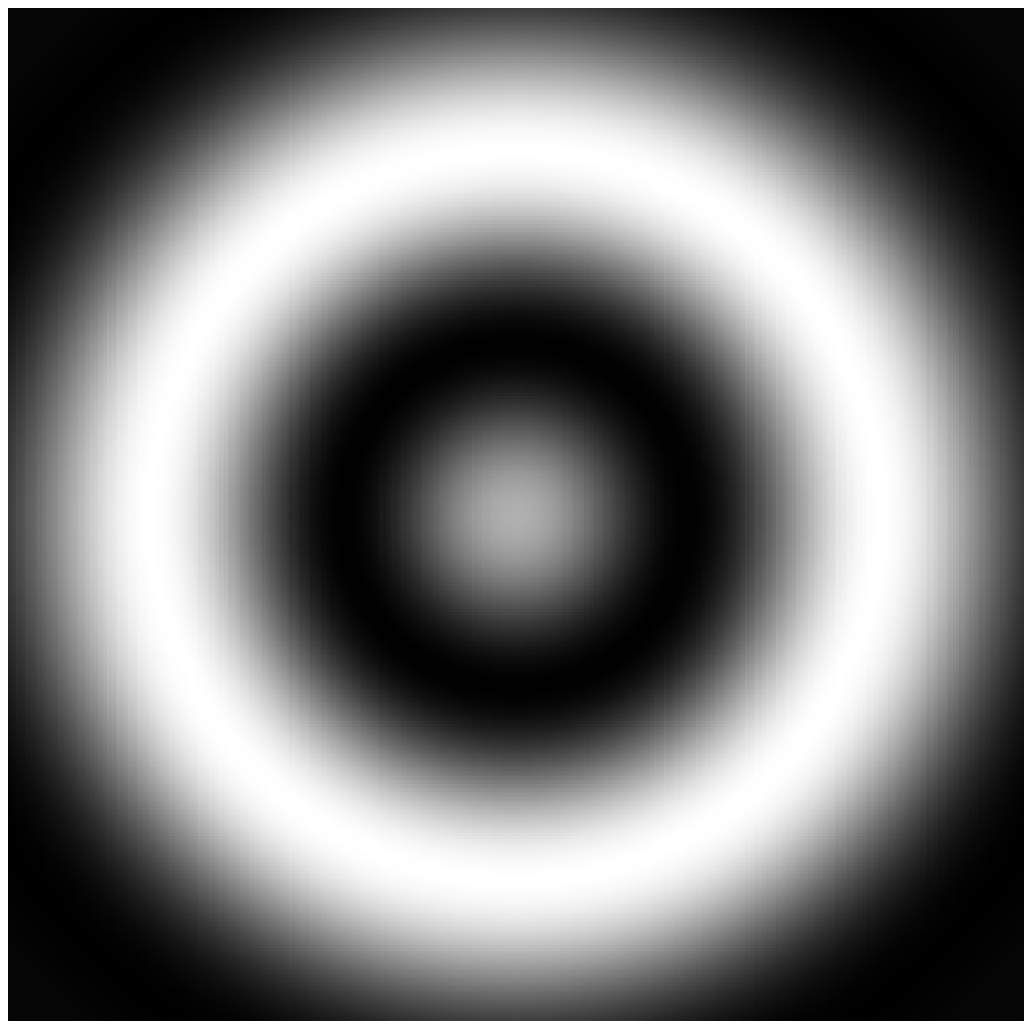}
\includegraphics[width=0.25\linewidth]{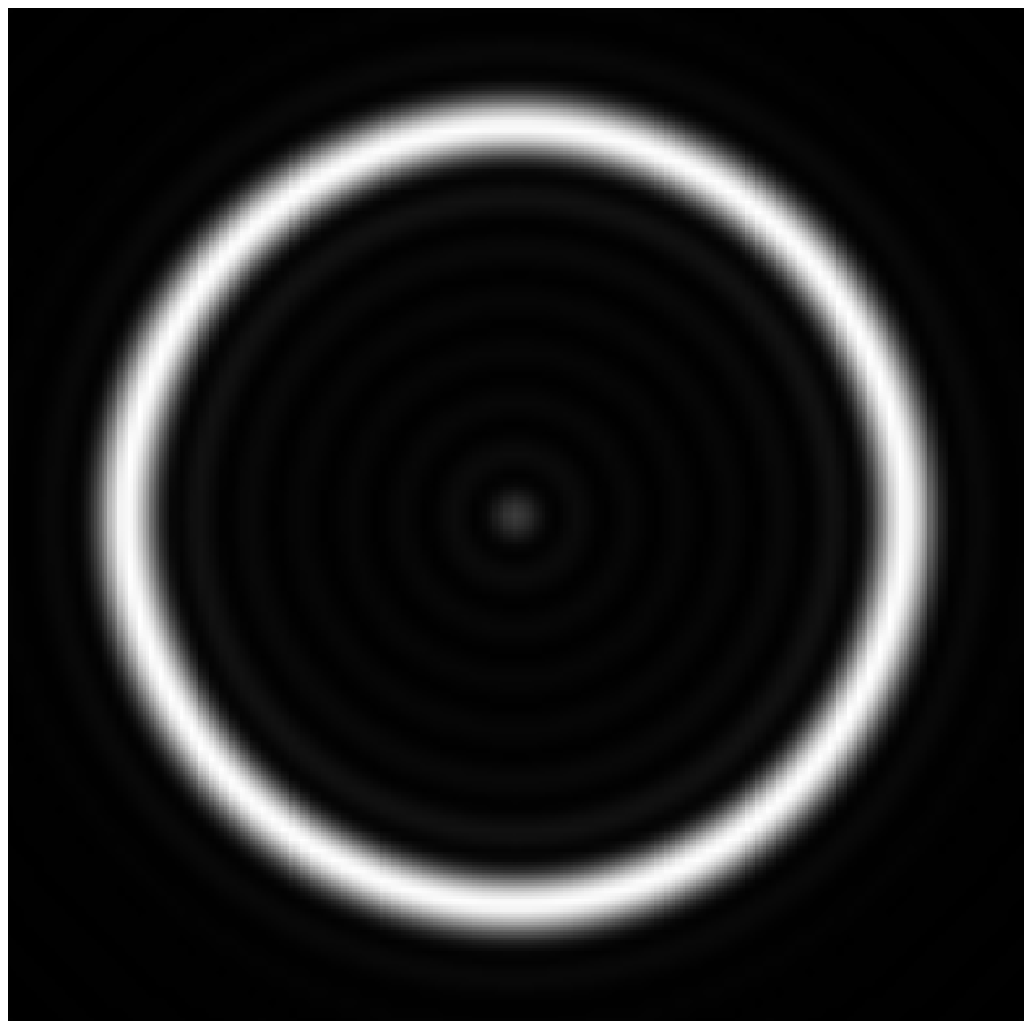}
\includegraphics[width=0.25\linewidth]{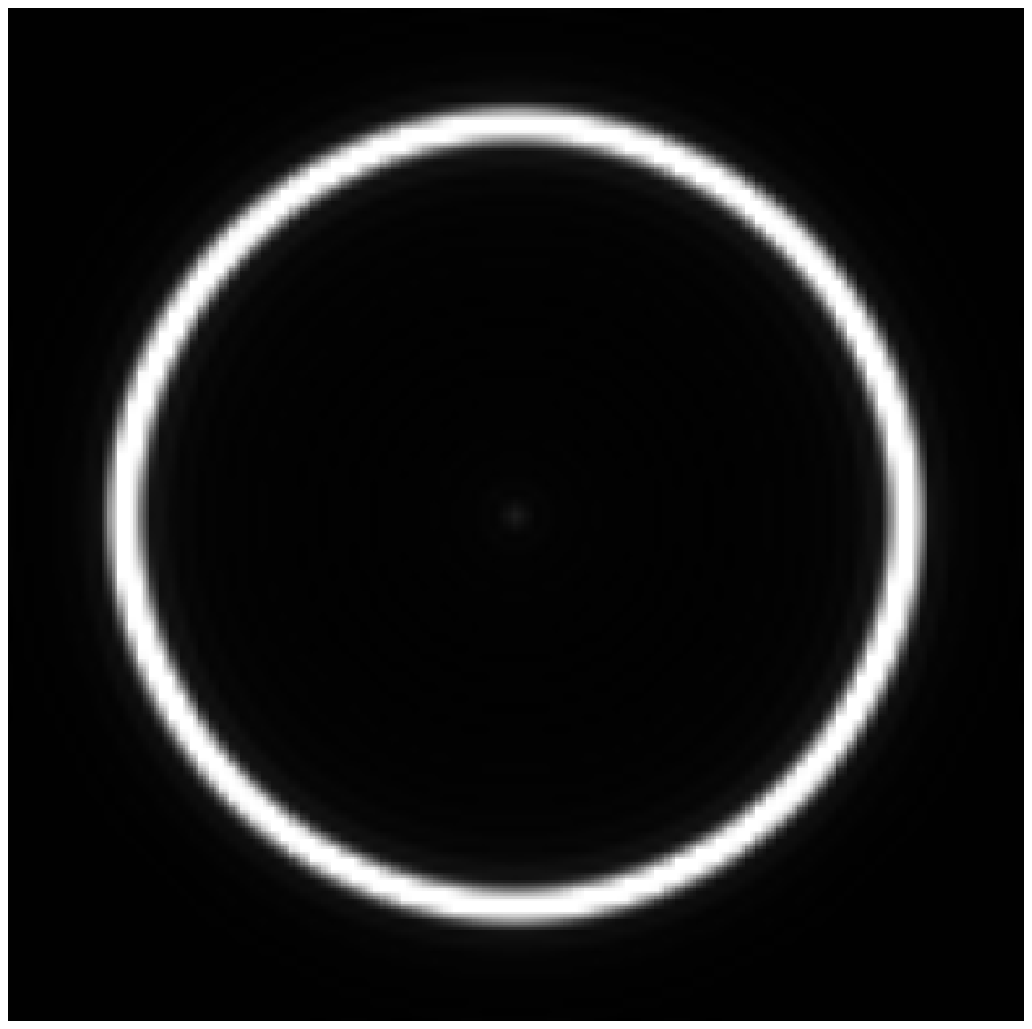}
\end{center}

\caption{\label{fig:ER-plot}Image formation in the sensor plane of an optical telescope with a 10~m focal length and an aperture of 25~cm (left), 1~m (center), vs. 2~m (right). The images depict the Einstein ring of a point source on the optical axis, as seen by the telescope. The image produced by a 25~cm aperture is dominated by the diffraction pattern of the telescope. Larger telescope apertures, though still diffraction-limited, offer sufficient resolution, e.g., for the use of a coronagraph to block out light from the Sun inside the Einstein ring.}
\end{figure}

One may show that expression (\ref{eq:amp-w-f2dp}) is always finite. In fact, even
when ${2r_g}/{\overline z}-{\rho_i^2}/{f^2}=0$, (\ref{eq:amp-w-f2dp}) remains finite and describes the Einstein ring as it seen is at the detector at the position given as
{}
\begin{eqnarray}
\rho_i=f\sqrt{\frac{2r_g}{\overline z}}.
  \label{eq:amp-det}
\end{eqnarray}
For the image of the Einstein ring to be resolved on the detector, the image size has to occupy several pixels (Fig.~\ref{fig:ER-plot}). Assuming that the pixel size is $\delta_p=10\,\mu$m, and the image occupies $n_0=10$ pixels, so that $\rho_i=n_0\delta_p$, such an imaging system would require a lens with the focal length of
{}
\begin{eqnarray}
f=n_0\delta_p\Big({\frac{\overline r}{2r_g}}\Big)^\frac{1}{2}=12.83\,{\rm m}\,\Big(\frac{n_0}{10}\Big)\Big(\frac{\delta_p}{10\,\mu{\rm m}}\Big)\Big({\frac{\overline z}{650\,{\rm AU}}}\Big)^\frac{1}{2}.
  \label{eq:amp-de2t}
\end{eqnarray}

We can also derive the power at the Einstein ring deposited on the detector. For this, in (\ref{eq:amp-w-f2dp}), we take a limit $\rho_i\rightarrow f\sqrt{{2r_g}/{\overline z}}$ and, relying on (\ref{eq:BF}), obtain
{}
\begin{eqnarray}
S_z(\rho_i^{\tt ER})=
\frac{c}{8\pi}\frac{E^{\tt s2}_0}{(\overline z+z_0)^2}\Big(\frac{kd^2}{8  f}\Big)^2
\,\frac{8\lambda {\overline z}}{\pi^2 d^2}.
  \label{eq:amp-w-f2dps}
\end{eqnarray}
Comparing this expression with (\ref{eq:amp-w-f2d2p}), we see that the light at the Einstein ring is amplified with the amplification factor given by the following expression:
{}
\begin{eqnarray}
\mu_{\tt det}=\frac{8\lambda {\overline z}}{\pi^2 d^2}=7.88\times 10^7\,\Big(\frac{\lambda}{1\,\mu{\rm m}}\Big)\Big(\frac{{\overline z}}{650\,{\rm AU}}\Big)\Big(\frac{1\,{\rm m}}{d}\Big)^2.
  \label{eq:amp-w-f2dps1}
\end{eqnarray}

We note that (\ref{eq:amp-w-f2dp}) is also finite when $|\rho_i|=0$ with  the corresponding value computed as
{}
\begin{eqnarray}
S_z(0)=
\frac{c}{8\pi}\frac{E^{\tt s 2}_0}{(\overline z+z_0)^2}\Big(\frac{kd^2}{8  f}\Big)^2
\mu_0\bigg(\frac{2J_1\Big({\textstyle\frac{1}{2}}kd \sqrt{\frac{2r_g}{\overline z}}\Big)}{{\textstyle\frac{1}{2}}kd \sqrt{\frac{2r_g}{\overline z}}}\bigg)^2.
  \label{eq:amp-w-f2dp7}
\end{eqnarray}
From (\ref{eq:amp-w-f2dp7}), using (\ref{eq:BF}), for $r_g\not=0$, the amplification factor at the center of the detector is evaluated to be
{}
\begin{eqnarray}
\mu^0_{\tt det}=\mu_0\bigg(\frac{2J_1\Big({\textstyle\frac{1}{2}}kd \sqrt{\frac{2r_g}{\overline z}}\Big)}{{\textstyle\frac{1}{2}}kd \sqrt{\frac{2r_g}{\overline z}}}\bigg)^2=\frac{16\lambda^2 \overline z}{\pi^2 d^3}\sqrt{\frac{\overline z}{2r_g}}\sin^2\Big[\frac{\pi d}{\lambda} \sqrt{\frac{2r_g}{\overline z}}-{\frac{\pi}{4}}\Big]=2.02\times 10^7\,\Big(\frac{\lambda}{1\,\mu{\rm m}}\Big)^2\Big(\frac{{\overline z}}{650\,{\rm AU}}\Big)^{\frac{3}{2}}\Big(\frac{1\,{\rm m}}{d}\Big)^3.
  \label{eq:amp-w-f2dps2}
\end{eqnarray}
For $r_g\rightarrow 0$,  the amplification factor $\mu^0_{\tt det}$ reduces to $\mu^0_{\tt det}=1$ and the result (\ref{eq:amp-w-f2dp7}) is equivalent to (\ref{eq:amp-w-f2d2p}) for $\rho_i=0$.

\section{Discussion and Conclusions}
\label{sec:disc}

Our main motivation for this paper was to study image formation by the SGL in the case of an extended source located at a large, but finite distance from the Sun. This is the situation that we encounter when considering the use of the SGL for multipixel imaging and spatially-resolved spectroscopy of exoplanets. It is also relevant to many scenarios that involve imaging of extended sources with microlensing techniques. Yet surprisingly, no theoretical description of this scenario could be found in the literature that we surveyed, especially from the wave-optical perspective.

We began by recalling results of our investigations were we studied the propagation of plane EM waves on the background of a gravitational monopole. We relied on a Mie theory that we developed \cite{Turyshev-Toth:2017} to account for the refractive properties of the gravitational field in the vicinity of the Sun. The resulting EM field is described in full by the Debye potential (\ref{eq:Pi-s_a+0}), which accounts for the fully-absorbing boundary conditions introduced at the solar surface.

A key step in the new derivation is the development of the radial function (\ref{eq:Fass*}) that accounts for the phase shift acquired along the path from the source to the image plane. We obtained such a function by using the WKB approximation (\ref{eq:R_solWKB+=_bar-imp}), which is typically used to solve similar problems in nuclear scattering. Using the approximate solution to the radial function, we developed the corresponding Debye potential and show that no EM field exists directly behind the Sun in the shadow region.

Using the Debye potential (\ref{eq:Pi-s_a+0}) and  the radial function (\ref{eq:Fass*}) in the region of geometric optics (see Sec.~\ref{sec:go-em-outside}) yields a solution for the EM wave propagating in this region. The EM field in the region of geometric optics can be described using incident and scattered waves, given by (\ref{eq:DB-t-pl=p10})--(\ref{eq:DB-t-pl=p20}) and (\ref{eq:scat-th-tot}), correspondingly. These solutions extend our previous results \cite{Turyshev-Toth:2017,Turyshev-Toth:2019} to the case of a source at finite distance.  A source that is positioned at a finite distance from the Sun with coordinates $({\vec b},r_0)$ can be dealt with by simply introducing a rotation by the angle $\beta=b/r_0$, in the plane defined by $\vec{b}$, thus defining an optical axis of the SGL for this particular point source. This optical axis connects the point source and the center of the Sun, extending towards the image plane. Rays of light envelop the entire solar circumference and propagate toward this optical axis, ultimately reaching it in the interference region (see Sec.~\ref{sec:IF-region}). An observer at this location would see a perfect Einstein ring formed around the Sun. A deviation from the optical axis results in breaking the ring into arcs of uneven brightness.

The most practically interesting solution was obtained for the EM field in the interference region (\ref{eq:DB-sol-rho}). In the case of a source at finite distance, for any given impact parameter, the focal point of the SGL, which is nominally given as $b^2/(2r_g)$, is shifted further out from the Sun by the extra distance $(b^2/2r_g)^2/z_0$ that must be accounted for in any SGL mission design and simulations.

We extend our approach to imaging of extended sources (see Sec.~\ref{sec:image_form}). For this we represent the surface of an extended source as a collection of point sources, each selecting its own optical axis while imaged via the SGL. We show that for each point in the source plane, there exists an optical axis that points to a specific point on the image plane, where the intensity of light from that point source is maximal. As the SGL's point-spread function is much broader than the classical Airy pattern, there will be light deposited at widely separated points on the image plane, albeit at weaker levels. A telescope of modest size, positioned in the image plane would see an Einstein ring containing light from the entire extended body. At each location on the image plane, such a ring is composed of the complete Einstein ring produced by light from the source location exactly opposite to the image plane location with respect to the center of the Sun, and partial arcs from all other points in the source plane. These rings and arcs together determine the power density of the signal received by a telescope in the image plane and, ultimately, the image formed on the telescope's optical sensor.
Our treatment leads to an expression for the surface power density in the image plane  (\ref{eq:power_dens}) that describes the total power received by an imaging telescope (\ref{eq:power_rec2}). The expressions obtained in this paper generalize previously known results to the case of a resolved extended source at a finite distance from the Sun.

The analysis can be extended to incorporate the optics of a telescope that would be positioned at the focal plane to collect light from the distant source. We presented an analytic derivation of the Einstein ring that appears due to a distant point source. This result is directly relevant to studying the required aperture of the telescope and the coronagraph that will be used to block out direct sunlight and much of the light from the solar corona, without obscuring light from the Einstein ring.

The next step is to analyze the signals received from realistic sources in our stellar neighborhood.  This work is ongoing and results, when available, will be reported elsewhere.

\begin{acknowledgments}
This work in part was performed at the Jet Propulsion Laboratory, California Institute of Technology, under a contract with the National Aeronautics and Space Administration.

\end{acknowledgments}


\appendix

\section{Solution for the radial equation in the WKB approximation}
\label{sec:rad_eq_wkb}

Here we focus on the equation for the radial function, $R_\ell$, used to identify the solution for the Debye potential (\ref{eq:Pi-s_a+0}). This equation has the following form (see Eqs. (25) and (F1) in \cite{Turyshev-Toth:2017}):
{}
\begin{eqnarray}
\frac{d^2 R_\ell}{d r^2}+\Big(k^2(1+\frac{2r_g}{r})+\frac{r_g}{r^3}-\frac{\ell(\ell+1)}{r^2}
\Big)R_\ell&=&0.
\label{eq:R-bar-k*2a}
\end{eqnarray}

Following an approach presented in \cite{Turyshev-Toth:2017}, we explore an approximate solution to (\ref{eq:R-bar-k*2a}) using the method of stationary phase (i.e., the Wentzel--Kramers--Brillouin, or WKB approximation \cite{Friedrich-Trost:2004}). As we are interested in the case when $k$ is rather large (for optical wavelengths $k=2\pi/\lambda=6.28\cdot10^6\,{\rm m}^{-1}$), we are looking for an asymptotic solution as $k\rightarrow\infty$.   In fact, we are looking for a solution  for $R_\ell$ in the form
{}
\begin{eqnarray}
R=\exp\Big[\int_{r_0}^r i \Big(k\alpha_{-1}(t)+\alpha_0(t)+k^{-1}\alpha_1(t)+...+k^{-n}\alpha_n(t)+...\Big)dt\Big].
\label{eq:R-exp-bar}
\end{eqnarray}
Defining $'= d/ d r$, with the help of a substitution of $R'/R=w$, for the function $w$ we obtain the following equation:
{}
\begin{eqnarray}
w'+w^2+k^2(1+\frac{2r_g}{r})+\frac{r_g}{r^3}-\frac{\alpha}{r^2}=0,
\label{eq:ricati_bar}
\end{eqnarray}
where $\alpha= \ell(\ell+1)$.
Using this substitution, up to the terms $\propto k^{-1}$, we have \cite{Turyshev-Toth:2017,Turyshev-Toth:2019}
{}
\begin{eqnarray}
w=i\Big(k\alpha_{-1}(r)+\alpha_0(r)+k^{-1}\alpha_1(r)+
...+
k^{-n}\alpha_n(r)+...\Big).~~
\label{eq:w_k_bar}
\end{eqnarray}
Substituting (\ref{eq:w_k_bar}) into (\ref{eq:ricati_bar}) and equating the terms with respect to the same powers of $k$, we get
{}
\begin{eqnarray}
\alpha^2_{-1}(r)=1+\frac{2r_g}{r}, \qquad i\alpha'_{-1}(r)-2\alpha_{-1}(r)\alpha_0(r)&=&0,\nonumber\\
i\alpha'_{0}(r)-\alpha^2_{0}(r)-2\alpha_{-1}(r)\alpha_{1}(r)+\frac{r_g}{r^3}-\frac{\alpha}{r^2}=0\qquad
 i\alpha'_{1}(r)-2\alpha_{-1}(r)\alpha_{2}(r)
-2\alpha_{0}(r)\alpha_{1}(r)&=&0.
\label{eq:series2_bar}
\end{eqnarray}
These equations, to the order of ${\cal O}(k^{-5}, r_g^2)$, may be solved as
{}
\begin{eqnarray}
\alpha_{-1}(r)&=&\pm(1+\frac{r_g}{r}), \qquad
\alpha_0(r)=-i\frac{r_g}{2r^2}, \qquad
\alpha_{1}(r)=\mp\frac{\alpha}{2r^2}(1-\frac{r_g}{r}),
...
\label{eq:series3_bar}
\end{eqnarray}
Note that the $\pm$ signs in these expressions are not independent; they all come from the solution for $\alpha_{-1}(r)$ in (\ref{eq:series3_bar}).

Substituting solution for $\alpha_{-1}(r)$ from (\ref{eq:series3_bar}) into (\ref{eq:R-exp-bar}), we have
{}
\begin{eqnarray}
S_{-1}(r)&=&\int_{r_0}^r \alpha_{-1}(\hat r)d\hat r=\pm\int_{r_0}^r (1+\frac{r_g}{\hat r})d\hat r.
\label{eq:S-1a}
\end{eqnarray}

We evaluate this integral along the part of a ray moving on the background of the solar gravitational field as given by the metric (\ref{eq:metric-gen})--(\ref{eq:pot_w_1**}).  With $\vec k$ being the unit vector in the gravitationally unperturbed direction of the light ray's propagation and with ${\vec x}_0$ being the initial position of the light ray, we introduce the parameter $\tau=\tau(t)$ as \cite{Turyshev-Toth:2017,Turyshev-Toth:2019}
{}
\begin{eqnarray}
\tau &=&({\vec k}\cdot {\vec x})=({\vec k}\cdot {\vec x}_{0})+c(t-t_0).
\label{eq:x-Newt*=}
\end{eqnarray}
Clearly, this quantity is negative all the way up to the closest approach where it changes the sign.

The trajectory of a light ray in these conditions was developed in \cite{Turyshev-Toth:2017,Turyshev-Toth:2019} and to the post-Newtonian level it given by
{}
\begin{eqnarray}
{\vec r} (\tau)&=&{\vec b}+{\vec k} \tau-r_g\Big({\vec k}\ln\frac{\tau+\sqrt{b^2+\tau^2}}{\tau_0+\sqrt{b^2+\tau^2_0}}+\frac{\vec b}{b^2}\big(\sqrt{b^2+\tau^2}+\tau-\sqrt{b^2+\tau^2_0}-\tau_0\big)\Big)+{\cal O}(r_g^2),
\label{eq:X-eq4*}
\end{eqnarray}
where $ {\vec b}=[[{\vec k}\times{\vec x}_0]\times{\vec k}]+{\cal O}(G)$ is the impact parameter of the unperturbed trajectory of the light ray. The vector ${\vec b}$ is directed from the origin of the coordinate system toward the point of the closest approach of the unperturbed path of light ray to that origin.
The radial distance specified by (\ref{eq:X-eq4*}) is computed to be
{}
\begin{eqnarray}
|{\vec r} (\tau)|&=&\sqrt{b^2+\tau^2}-r_g\Big(1+\frac{\tau}{\sqrt{b^2+\tau^2}}\big(1+\ln\frac{\tau+\sqrt{b^2+\tau^2}}{\tau_0+\sqrt{b^2+\tau^2_0}}\big)-\frac{\sqrt{b^2+\tau^2_0}+\tau_0}{\sqrt{b^2+\tau^2}}\Big)+{\cal O}(r_g^2).
\label{eq:X-eq4*-rad}
\end{eqnarray}
At the closest approach, where $\tau=0$, the expression (\ref{eq:X-eq4*-rad}) takes the form:
{}
\begin{eqnarray}
r_{\tt c.a.}=|{\vec r} (0)|&=&b-r_g+{\cal O}(r_g^2,r_gb/\tau_0).
\label{eq:X-eq4*-rad-cla}
\end{eqnarray}

In the case of a gravitational field produced by monopole, the motion is constrained to a plane $\phi=\phi_0$.
Considering the motion within that plane, and using the expressions above, we evaluate the integral (\ref{eq:S-1a}) as
{}
\begin{eqnarray}
S_{-1}(r)&=&\pm\Big(\int_{r_{\tt c.a.}}^r \hskip -3pt \big(1+\frac{r_g}{\hat r}\big)d\hat r+\int^{r_0}_{r_{\tt c.a.}} \hskip -3pt\big(1+\frac{r_g}{\hat r}\big)d \hat r\Big) =\pm \Big(r+r_0-2b+r_g\ln \frac{rr_0e^2}{b^2}\Big).~~~
\label{eq:S-1}
\end{eqnarray}
This expression describes the geometric phase delay experienced by an EM wave as it travels from the source at a large but finite distance from the Sun, $r_0$, to the image plane in the focal region of the SGL located at heliocentric distance $r$. Similarly, we integrate the remaining equations (\ref{eq:series3_bar}):
{}
\begin{eqnarray}
S_{0}(r)&=&\int_{r_0}^r \alpha_{0}(\hat r)d\hat r=-i\int_{r_0}^r \frac{r_g}{2\hat r^2}d\hat r=i\Big(\frac{r_g}{2r}+\frac{r_g}{2r_0}-\frac{1}{b}\Big),
\label{eq:S-0}\\
S_1(r)&=&\int_{r_0}^r \alpha_{1}(\hat r)d\hat r=\mp\frac{\alpha}{2}\int_{r_0}^r \frac{d\hat r}{\hat r^2}(1-\frac{r_g}{\hat r})=\pm\Big(\frac{\alpha}{2r}\Big(1-\frac{r_g}{2r}\Big)+\frac{\alpha}{2r_0}\Big(1-\frac{r_g}{2r_0}\Big)-\frac{\alpha}{b}\Big(1+\frac{r_g}{2b}\Big)\Big).
\label{eq:S1}
\end{eqnarray}

As a result, an approximate solution for the partial radial function $R_\ell$ to ${\cal O}\big((kr)^{-2},r_g^2\big)$  is given as \cite{Turyshev-Toth:2017,Turyshev-Toth:2019}
{}
\begin{eqnarray}
R_\ell(r)&=& c_\ell \exp\Big[i\big(kS_{-1}(r)+S_0(r)+k^{-1}S_1(r)
\big)\Big]+
d_\ell \exp\Big[-i\big(kS_{-1}(r)+S_0(r)+k^{-1}S_1(r)
\big)\Big],
\label{eq:R_solWKB+=_bar+}
\end{eqnarray}
where $c_\ell$ and $d_\ell$ are arbitrary constants. Substituting  (\ref{eq:S-1})--(\ref{eq:S1}) in (\ref{eq:R_solWKB+=_bar+}), we obtain the following solution for $R_\ell$:
{}
\begin{eqnarray}
R_\ell(r)&=&
c_\ell \,\exp\Big[i\Big(k(r+r_0+r_g\ln4k^2rr_0)+\frac{\ell(\ell+1)}{2k}\Big(\frac{1}{r}+\frac{1}{r_0}\Big)
\Big)\Big]+\nonumber\\
&&+\,d_\ell\,
\exp\Big[-i\Big(k(r+r_0+r_g\ln4k^2rr_0)+\frac{\ell(\ell+1)}{2k}\Big(\frac{1}{r}+\frac{1}{r_0}\Big)\Big)\Big]+
{\cal O}\big((kr)^{-2},r_g^2\big),
\label{eq:R_solWKB+=_bar}
\end{eqnarray}
where $c_\ell$ and $d_\ell$  now account for all the integration constants in (\ref{eq:S-1})--(\ref{eq:S1}).  In addition, we omitted the terms $\propto r_g/r$ in (\ref{eq:S-1})--(\ref{eq:S1}) as their contribution to the amplitude of the EM wave is negligibly small (see \cite{Turyshev-Toth:2017,Turyshev-Toth:2019} for details).

We may further improve the asymptotic expression for $R_\ell$ from (\ref{eq:R_solWKB+=_bar}) by accounting for the Coulomb phase shifts, $\sigma_\ell$, that an EM wave experiences after passing by a gravitating monopole, which can be done by simply redefining the constants $c_\ell$ and $d_\ell$ yet again (see discussion in \cite{Turyshev-Toth:2017}) as
{}
\begin{eqnarray}
c_\ell\rightarrow c_\ell \exp\big[{i(\sigma_\ell-\frac{\pi \ell}{2})}\big],\qquad\qquad d_\ell\rightarrow d_\ell \exp\big[{-i(\sigma_\ell-\frac{\pi \ell}{2})}\big].
\label{eq:cd_ell}
\end{eqnarray}

As a result, the expression for the asymptotic behavior of the partial radial function $R_\ell$ takes the form \cite{Turyshev-Toth:2017,Turyshev-Toth:2019}:
\begin{eqnarray}
R_\ell(r)&=&
c_\ell \,\exp\Big[i\Big(k(r+r_0+r_g\ln 4k^2rr_0)+\frac{\ell(\ell+1)}{2k}\Big(\frac{1}{r}+\frac{1}{r_0}\Big)+\sigma_\ell-\frac{\pi \ell}{2}\Big)\Big]+\nonumber\\
&+&
d_\ell\,\exp\Big[-i\Big(k(r+r_0+r_g\ln 4k^2rr_0)+\frac{\ell(\ell+1)}{2k}\Big(\frac{1}{r}+\frac{1}{r_0}\Big)+\sigma_\ell-\frac{\pi \ell}{2}\Big)\Big]
+{\cal O}\big((kr)^{-2}, r_g^2\big).~~~~~
\label{eq:R_solWKB+=_bar-imp}
\end{eqnarray}

In \cite{Turyshev-Toth:2017} the asymptotic behavior of the Coulomb function was obtained for very large distances from the turning point, $r\gg r_{\tt t}$; the solution (\ref{eq:R_solWKB+=_bar-imp}) improves it further by extending the argument of these functions to shorter distances, closer to the turning point while also presenting its explicit dependent on the distance to the source.
\end{document}